\documentclass[11pt]{article}
\usepackage[USenglish]{babel}
\usepackage{a4wide}
\usepackage{epsfig}
\usepackage[all]{xy}
\usepackage{amssymb}
\usepackage{amsmath}
\usepackage{latexsym}
\usepackage{pifont}
\usepackage{pdfsync}
\usepackage{enumerate}
\usepackage[parfill]{parskip}
\usepackage{xr}
\usepackage{euler}
\usepackage[utf8]{inputenc}  
\title{Algebraic Properties of Stochastic Effectivity Functions}
\author{Ernst-Erich Doberkat\thanks{Funded in part by Deutsche Forschungsgemeinschaft, grant DO 263/12-1, \emph{Koalgebraische Eigenschaften stochastischer Relationen.}}\\
Technische Universität Dortmund\\
\texttt{ernst-erich.doberkat@udo.edu}\\[6pt]
\textit{Dedicated to Professor Prakash Panangaden on the occasion of his 60th birthday.}}
\date{}
\newcommand{\labelImpl}[2]{\ensuremath{\ref{#1}~\Rightarrow~\ref{#2}}}

\newcommand{\Inv}[2]{\ensuremath{{\mathcal INV}\left(#1, #2\right)}}
\newcommand{\Klasse}[2]{\left[#1\right]_{#2}}
\newcommand{\Faktor}[2]{{#1}/{#2}}
\newcommand{\fMap}[1]{\eta_{#1}}
\newcommand{\Bild}[2]{{#1}\left[#2\right]}
\newcommand{\InvBild}[2]{\Bild{#1^{-1}}{#2}}
\newcommand{\Kern}[1]{\mathsf{ker}\left(#1\right)}
\newcommand{\Folge}[1]{(#1_n)_{n \in \Nat}}

\newcommand{\spaceFont}[1]{\mathfrak{#1}}

\newcommand{\SubProb}[1]{\spaceFont{S}\left(#1\right)}

\newcommand{\SubProbSenza}{\spaceFont{S}}

\newcommand{\PowerSet}[1]{\ensuremath{\mathcal{P}\left(#1\right)}}
\newcommand{\PowerSenza}{\ensuremath{\mathcal{P}}}

\newcommand{\Borel}[1]{\ensuremath{{\mathcal B}(#1)}}

\edef\LinkeKlammer{\lbrack\!\lbrack}
\edef\RechteKlammer{\rbrack\!\rbrack}
\newcommand{\Gilt}[1][\phi]{\ensuremath{\LinkeKlammer#1\RechteKlammer}}

\newcommand{\Trans}{\rightsquigarrow}

\newtheorem{definition}{Definition}[section]
\newcommand{\BeginDefinition}[1]{%
  \begin{definition}\label{#1}
}
\newcommand{\EndDefinition}{\end{definition}}

\newtheorem{example}[definition]{Example}
\newcommand{\BeginExample}[1]{%
  \begin{example}\label{#1}\rm
}
\newcommand{\EndExample}{--- \end{example}}

\newtheorem{observation}[definition]{Observation}
\newcommand{\BeginObservation}[1]{
  \begin{observation}\label{#1}\rm
}
\newcommand{\EndObservation}{--- \end{observation}}

\newtheorem{theorem}[definition]{Theorem}
\newcommand{\BeginTheorem}[1]{%
  \begin{theorem}\label{#1}
}
\newcommand{\EndTheorem}{\end{theorem}}

\newtheorem{corollary}[definition]{Corollary}
\newcommand{\BeginCorollary}[1]{
  \begin{corollary}\label{#1}
}

\newtheorem{proposition}[definition]{Proposition}
\newcommand{\BeginProposition}[1]{%
  \begin{proposition}\label{#1}
}
\newcommand{\EndProposition}{\end{proposition}}

\newcommand{\EndCorollary}{\end{corollary}}
\newtheorem{lemma}[definition]{Lemma}
\newcommand{\BeginLemma}[1]{%
  \begin{lemma}\label{#1}
}
\newcommand{\EndLemma}{\end{lemma}}

\newtheorem{claim}{Claim}
\newcommand{\BeginClaim}[1]{%
  \begin{claim}\label{#1}
}
\newcommand{\EndClaim}{\end{claim}}

\newenvironment{proof}{\textbf{Proof\ }}{\ensuremath{\QED}}
\newcommand{\BeginProof}{\begin{proof}}
\newcommand{\EndProof}{\end{proof}}

\newenvironment{remark}{\textbf{Remark:\ }}{}
\newcommand{\BeginRemark}{\begin{remark}}
\newcommand{\EndRemark}{\QED\end{remark}}
\newcommand{\QED}{%
\ensuremath{\dashv}
}

\newcommand{\Real}{\mathbb{R}}
\newcommand{\pReal}{\mathbb{R}_{+}}
\newcommand{\Nat}{\mathbb{N}}
\newcommand{\Rational}{\mathbb{Q}}

\def\cal{\mathcal}
\def\mathcal{\mathscr}
\renewcommand{\Borel}[1]{\ensuremath{{\spaceFont B}(#1)}}
\renewcommand{\spaceFont}{\mathcal}
\def\funktorFont{\mathbb}
\def\VauSenza{\ensuremath{\funktorFont{V}}}
\def\WeSenza{\ensuremath{\funktorFont{W}}}
\newcommand{\Vau}[1]{\ensuremath{\ensuremath{\VauSenza(#1)}}}
\newcommand{\We}[1]{\ensuremath{\ensuremath{\WeSenza(#1)}}}

\newcommand{\basS}[3]{\bas{#2}{#3 \ast #1}}
\renewcommand{\basS}[3]{\bas{#2}{#3\,\pmb{\mid}\,#1}}
\renewcommand{\basS}[3]{{\boldsymbol{\beta}}_{#1}(#2, #3)}

\newcommand{\isEquiv}[3]{\ensuremath{{#1}\ {#3}\ {#2}}}
\renewcommand{\Inv}[2]{\ensuremath{\pmb{\Sigma}_{#1}(#2)}}
\renewcommand{\Inv}[2]{\ensuremath{{\Sigma}(#1, #2)}}
\renewcommand{\SubProbSenza}{\funktorFont{S}}
\renewcommand{\SubProb}[1]{\SubProbSenza(#1)}

\renewcommand{\Gilt}[2][\phi]{\LinkeKlammer#1\RechteKlammer_{#2}}

\def\phi{\varphi}
\def\theta{\vartheta}

\def\QED{\ensuremath{\hfill\Box}}

\makeatletter
\newcommand{\@testOp}[2]{#1#2}
\newcommand{\posTest}[1][\phi]{\@testOp{#1}{?}}
\newcommand{\negTest}[1][\phi]{\@testOp{#1}{\text{\textquestiondown}}}
\makeatother
\def\eTrans{\pmb{\twoheadrightarrow}}
\makeatletter
\newcommand{\@BSS}[1]{\Borel{{#1}\otimes[0, 1]}}
\newcommand{\@boss}[1]{\Borel{\SubProb{#1}}}
\renewcommand{\@boss}[1]{w({\mathcal #1})}
\def\BSO{\@BSS{S}}
\def\BSOp{\@BSS{S'}}
\def\BSSO{\@BSS{\SubProb{S}}}
\def\BTO{\@BSS{T}}
\def\BSTO{\@BSS{\SubProb{T}}}
\def\BSTOp{\@BSS{\SubProb{T'}}}
\def\boss{\@boss{S}}
\def\bost{\@boss{T}}
\makeatother
\newcommand{\bSS}[1]{{#1}\otimes\Borel{[0, 1]}}
\newcommand{\hut}[1]{\ensuremath{{#1}^{\flat}}}
\newcommand{\pP}[1][p]{\mathfrak{#1}}
\def\cG#1{\ensuremath{\mathsf{#1}}}
\renewcommand{\Bild}[2]{{#1}\bigl[#2\bigr]}

\usepackage{fancyhdr}
\begin{document}
\maketitle
\begin{abstract}
  Effectivity functions are the basic formalism for
  investigating the semantics game logic. We discuss algebraic properties
  of stochastic effectivity functions, in particular the relationship to
  stochastic relations, morphisms and congruences are defined, and the
  relationship of abstract logical equivalence and behavioral equivalence is
  investigated. 
\end{abstract}


\section{Introduction}
\label{sec:introduction}

This paper investigates some algebraic properties of stochastic
effectivity functions, which have been used as the basic formalism for
interpreting game logic, and which show some interesting relationships
to non-deterministic labelled Markov processes and to stochastic
relations, i.e., to Markov transition systems. Before entering into
the discussion of the stochastic branch of this family of functions,
it is interesting and illuminating to have a look at the evolution of these functions,
so let us start with some historical remarks.

\paragraph{Historical Remarks.}
\label{sec:historic-remarks}

Effectivity functions were first systematically investigated in the
area of Social Choice, and here in particular for the modeling of
voting systems~\cite[Chapter 7.2]{Moulin-SocialChoice}. Moulin models
the outcome of cooperative voting through a binary relation
$\mathbf{E} \subseteq \PowerSet{N}\times\PowerSet{A}$ between
coalitions of voters and subsets of outcomes, where a coalition is
just a subset of the entire population. Here $N$ is the set of voters,
$A$ the set of outcomes, and $\PowerSenza$ denotes the power set. If
$T\ \mathbf{E}\ B$, coalition $T$ is said to be \emph{effective for}
the subset $B$ of outcomes, thus coalition $T$ can force an outcome in
$B$. Among others, Moulin postulates that if $T$ is effective for $B$,
then $N\setminus T$ must not be effective for $S\setminus B$. Some
examples (unanimity with status quo, veto functions) illustrate the
approach.  

Generalizing this in their work on Social Choice, Abdou and
Keiding~\cite{Abdou-Keiding} define effectivity functions as special
cases of conditional game forms. Such a \emph{conditional game form}
is a map $E:
\PowerSet{N}\to\PowerSet{\mathcal{A}\setminus\{\emptyset\}}$, where
$N$ and $A$ are as above, $\mathcal{A}$ is a subset of $\PowerSet{A}$
with $\emptyset\in\mathcal{A}, A\in\mathcal{A}$. The family of all
closed sets in a topological space or of all measurable subsets in a
measurable space are mentioned as examples. If $T$ is a coalition,
$B\in E(T)$ models that coalition $T$ can force an outcome in
$B$. Thus the notion of effectivity is the same as in Moulin's
proposal. Based on this, a first simple and general form of
effectivity function is defined. A conditional game form $E:
\PowerSet{N}\to\PowerSet{\mathcal{A}\setminus\{\emptyset\}}$ is called
an \emph{effectivity function} iff $A\in E(T)$ for all non-empty
coalitions $T$, and if $E(N) =
\mathcal{A}\setminus\{\emptyset\}$. Thus a non-empty coalition can
achieve something, and the community $N$ has all options to choose
from.

The functions discussed so far do not make an assumption on
monotonicity; for the purposes of the present paper, however, monotone
effectivity functions are of interest. This property appears to be natural,
given the interpretation of effectivity functions: If an outcome in $B$
can be forced by coalition $T$, then an outcome in each super set $B' \supseteq B$
can be forced by this coalition as well; if this is true for all
coalitions, function $E$ is called
\emph{monotone}.  

The neighborhood relations used here are taken from the minimal models
discussed in modal logics~\cite[Chapter 7.1]{Chellas}, serving as
basic mechanism for models which are more general than Kripke models. The association of the effectivity
functions sketched here to a very similar notion investigated in
economics is discussed in the survey paper~\cite[Section
2.3]{Hoe-Pauly}.

\paragraph{Game Logic.}
\label{sec:game-logic}

{\def\widetilde{\breve}
Parikh~\cite{Parikh-Games1985}, and later Pauly~\cite{Pauly-CWI}
propose interpreting game logic through a neighborhood model. Assign
to each primitive game $g$ and each player $\{1, 2\}$ a neighborhood
relation $N_g^{(i)} \subseteq S \times \PowerSet{S}\ (i = 1, 2)$ with
the understanding that $s N_g^{(i)} X$ indicates player $i$ having a
strategy in state $s$ to force a state in $X \subseteq S$. Here $S$ is
the set of states over which the game is interpreted. The fact that $s
N_g^{(i)} X$ is sometimes described by saying that player $i$ is
effective for $X$ (with game $g$ in state $s$). It is desirable that
$s N_g^{(i)} X$ and $X \subseteq X'$ imply $s N_g^{(i)} X'$ for all
states $s$. We assume in addition that the game is \emph{determined}, i.e., that
exactly one of the players has a winning strategy. Thus $X \subseteq
S$ is effective for player $1$ in state $s$ if and only if $S
\setminus X$ is not effective for player $2$ in that
state. Consequently,
\begin{equation}
\label{intro:det}
s N_g^{(2)} X \Leftrightarrow\neg(s N_g^{(1)} S\setminus X), 
\end{equation}
which in turn implies that we only have to cater for player~$1$. We
will omit the superscript from the neighborhood relation $N_g$. Define
a map $S \to \PowerSet{\PowerSet{S}} $, again denoted by $N_{g}$, upon setting $ N_g(s)
:= \{X \subseteq S \mid s N_g X\}, $ then $N_g(s)$ is an upper closed
subset of $\PowerSet{S}$ for all $s \in S$ from which relation $N_g$
can be recovered. This function is called the \emph{effectivity
  function} associated with relation $N_g$. From $N_g$ another map $
\widetilde{N}_g: \PowerSet{S}\to\PowerSet{S} $ is obtained upon
setting 
\begin{equation*}
\widetilde{N}_g(A) := \{s \in S \mid s N_g A\} = \{s \in S
\mid A \in N_g(s)\}.
\end{equation*}

Thus state $s$ is an element of
$\widetilde{N}_g(A)$ iff the first player has a strategy force the
outcome in $A$ when playing $g$ in $s$. The operations on games can be
taken care of through this family of maps, e.g., one sets recursively
for the first player
\begin{align}
\label{eq:1}\widetilde{N}_{g_1\cup g_2}(A) & := \widetilde{N}_{g_1}(A)\cup \widetilde{N}_{g_2}(A),\\
\label{eq:5}\widetilde{N}_{g_1;g_2}(A) & := (\widetilde{N}_{g_1}\circ \widetilde{N}_{g_2})(A),\\
\label{eq:6}\widetilde{N}_{g^*} & := \bigcup_{n \geq 0} \widetilde{N}_{g^n}(A),
\end{align}
with $g_1\cup g_2$ denoting the game in which the first player chooses from games $g_1$,
$g_2$, the game $g_1;g_2$ plays $g_1$ first, then $g_2$, and $g^*$ is
the indefinite iteration of game $g$. This refers only to player~$1$,
player~$2$ is accommodated through $A \mapsto S\setminus
\widetilde{N}_g(S \setminus A)$ by~(\ref{intro:det}), since the game
is determined. Pauly~\cite[Section 6.3]{Pauly-CWI} discusses the important point of
determinacy of games and relates it briefly to the discussion in set
theory~\cite[Section 33]{Jech},~\cite[Section 12.3]{Jech-Choice}
or~\cite[Section 20]{Kechris}. 

The maps $\widetilde{N}_g$ serve in Parikhs's original paper as a
basis for defining the semantics of game logic. It turns out to be
convenient for the present paper to go back, and to use effectivity
functions as maps to upper closed subsets. When interpreting game
logic probabilistically, however, constructions~(\ref{eq:1})
--~(\ref{eq:6}) are fairly meaningless, because one cannot talk about, e.g., 
the union of two probabilities. Hence another path had to be
travelled, which may be seen as a technical generalization of the
proposal~\cite{EED-PDL-TR} for interpreting propositional dynamic
logic.

\paragraph{Requirements.}
\label{sec:requirements}

The paper~\cite{EED-GameLogic-TR} proposes an approach through
stochastic effectivity functions. This variant assigns to each state
sets of probability distributions, to be specific, each state is
assigned an upward closed set of measurable sets of these
distributions. This property is inherited from the effectivity
functions discussed above. 

But let us have a look at other requirements for a stochastic
effectivity function. They should generalize stochastic relations in
the sense that each stochastic relation generates a stochastic
effectivity function in a fairly natural way; this requirement will
make it possible to consider the interpretation of modal logics
through stochastic Kripke models as interpretations through stochastic
effectivity functions, hereby enabling us to compare ---~or contrast,
as the case may be~--- results for these formalisms. We want also to
state that morphisms for stochastic relations are morphisms for the
associated effectivity functions. In addition, we want to have some
notion of measurability for these functions. Consequently, the set of all
distributions has to be made a measurable space. This raises
immediately the question of measurability of the effectivity function
proper, which would require a measurable structure on the upward
closed subsets of the measurable sets of the set of distributions over
the state space.

The design of non-deterministic labelled Markov processes proceeds
along this path: the set of measurable subsets of the measures on a
measurable space is equipped with a $\sigma$-algebra (akin to defining
a topology on the space of all closed subsets of a topological space),
and measurability is defined through this
$\sigma$-algebra~\cite{Terraf-Bisim-MSCS}, the \emph{hit
$\sigma$-algebra}. It turns out that one of the important issues here
is composability of effectivity functions for the purpose of catering for the
composition of games (or, more general, of actions). This is difficult
to achieve with the concept of hit measurability, and it will be taken
care of by the concept of t-measurability, which is introduced below.

Let us briefly discuss the issue of composability. Bind actions
$\gamma$ and $\delta$ to effectivity functions $P_{\gamma}$ and
$P_{\delta}$; we want to model action $\gamma;\delta$, the
sequential execution of $\gamma$ and $\delta$. Given a measurable set
$A$, the sets $\beta(A, r) := \{\mu \mid \mu(A) > r\}$ collect all the
evaluations for $A$, where $r\in[0, 1]$,so $\mu\in\beta(A, r)$ means
that event $A$ happens with probability greater than $r$, if the
probability is governed by $\mu$. Suppose that we know the set
$Q_{\delta}(A, r)$ of all states in which an evaluation of the
measurable set $A$ can be achieved through action $\delta$. We pick
all state distributions the expectation of which over $Q_{\delta}(A,
r)$ is greater than a threshold value $q$, hence we look at the set
\begin{equation}
\label{req_eq:7}
  \{\nu \mid \int_{0}^{1}\nu(Q_{\delta}(A, r))>q\}
\end{equation}
If this set is an element of $P_{\gamma}(s)$, then we say that the
composed action $\gamma;\delta$ can achieve a distribution which
evaluates at event $A$ with a probability greater than $q$. 

In order to get this slightly involved machinery going, however, we
have to make sure that the set described in~(\ref{req_eq:7}) is
actually a measurable of measures. Thus we need a suitable concept of
measurability, which we call \emph{t-measurability} and which is formulated
below.  
}

\paragraph{Overview.}
\label{sec:overview}

This paper discusses stochastic effectivity functions from a general
point of view, pointing out some algebraic properties which might be
helpful when using these functions for interpreting logics. We
introduce these functions after a careful discussion of the underlying
concept of measurability, and after comparing this concept to the one
proposed with a model for non-deterministic randomness. The
relationship to stochastic relations, or, as they are called
sometimes, sub probabilistic Markov kernels is discussed at length,
and it is shown that stochastic effectivity functions form
---~together with non-deterministic Markov processes as their close
cousins~--- a true generalization of these kernels. This applies as
well to morphisms and, in consequence, to congruences, which are
introduced and discussed. When interpreting logics through a
particular class of models, the question of expressivity becomes
important, the main issues being behavioral and logical equivalence,
with bisimilarity being another main actor in this play. We discuss
the question of behavioral and logical equivalence in the present
paper, formulating these properties without having to go back to an
underlying logic; this is done as in~\cite{EED-CongBisim} in purely
algebraic terms through morphisms and congruences. We give criteria
under which logically equivalent effectivity functions are
behaviorally equivalent.

Before we set out to discuss all these properties, we remind the
reader of some properties of measurable spaces, providing some tools
on the way. We also investigate some interesting properties of final
measurable surjections; they turn out to be helpful when it comes to
scrutinize $\sigma$-algebras of invariant sets and spaces of
subprobabilities. These maps turn out to be a very convenient means of
transporting measures faithfully, which becomes of interest when investigating
congruences.


\section{Basic Definitions}
\label{sec:basic-definitions}

The reader is briefly reminded of some notions and constructions from measure theory,
including the famous $\pi$-$\lambda$-Theorem and Choquet's
representation of integrals as areas; these two tools are used all
over the paper. We also introduce sets which are invariant under an equivalence
relation together with the corresponding $\sigma$-algebra of invariant
sets. As a special case of equivalence relations we introduce tame
relations, which are equivalence relations that are compatible with
quantitative measurements.

\subsection{Measurability}
First we fix some notations. A measurable space $(S, {\cal S})$ is a set $S$ with a
$\sigma$-algebra ${\cal S}$, i.e., ${\cal S}\subseteq\PowerSet{S}$ is
a Boolean algebra which is closed under countable unions. Here
$\PowerSet{S}$ is the power set of $S$. Given ${\cal S}_{0}\subseteq\PowerSet{S}$,
denote by 
\begin{equation*}
  \sigma({\cal S}_{0}) := \bigcap\{{\cal T} \mid {\cal
    S}_{0}\subseteq{\cal T}, {\cal T}\text{ is a $\sigma$-algebra}\}
\end{equation*}
the smallest $\sigma$-algebra containing ${\cal S}_{0}$ (the set for
which the intersection is constructed is not empty, since it contains
$\PowerSet{S}$). If $S$ is a topological space with topology $\tau$,
then the elements of $\sigma(\tau)$ are called the \emph{Borel sets}
of $S$; the $\sigma$-algebra $\sigma(\tau)$ is usually denoted by
$\Borel{S}$.

Given two measurable spaces $(S, {\cal S})$ and $(T,
{\cal T})$, the product space $(S\times T, {\cal S}\otimes{\cal T})$
has the Cartesian product $S\times T$ as a carrier set, the product
$\sigma$-algebra 
\begin{equation*}
  {\cal S}\otimes{\cal T} := \sigma(\{A\times B \mid A\in{\cal S}, B\in{\cal T}\})
\end{equation*}
is the smallest $\sigma$-algebra on $S\times T$ which contains all
measurable rectangles $A\times B$ with $A\in {\cal S}$ and $B\in{\cal
  T}$. Define for $E\subseteq S\times T$ 
\begin{align*}
  E^{s} & := \{t\in T\mid \langle s, t\rangle\in E\}&&\text{ (vertical
    cut),}\\
E_{t} & := \{s\in S \mid \langle s, t\rangle \in E\}&&\text{ (horizontal
    cut),}
\end{align*}
then $E^{s}\in{\cal T}$ for all $s\in S$, and $E_{t}\in{\cal S}$ for
all\footnote{The
  notation of indicating the horizontal cut through an index conflicts
  with indexing, but it is customary, so we will be careful to make
  sure which meaning we have in mind.} $t\in T$, provided $E\in{\cal S}\otimes{\cal T}$. The converse does not hold: Let
$S$ have a cardinality strictly larger than that of the continuum, then
$\PowerSet{S}\otimes\PowerSet{S}$ is a proper subset of
$\PowerSet{S\times S}$, because $\Delta := \{\langle s,
s\rangle \mid s\in S\}\not\in \PowerSet{S}\otimes\PowerSet{S}$~\cite[Exercise 21.20]{Hewitt-Stromberg}. On the other hand,
$\Delta_{s} = \{s\} = \Delta^{s}\in\PowerSet{S}$ for all $s\in S$,
so we cannot conclude that a set is product measurable, provided all its cuts are measurable. 

If $(S, \tau)$ and $(T, \theta)$ are topological spaces, then the Borel sets
$\Borel{\tau\times\theta}$ of the product topology may properly
contain the product $\Borel{\tau}\otimes\Borel{\theta}$. If, however,
both spaces are Hausdorff and $\theta$ has a countable basis, then
$\Borel{\tau\times\theta} =
\Borel{\tau}\otimes\Borel{\theta}$~\cite[Lemma 6.4.2]{Bogachev}. In
particular, the Borel sets of the product of two Polish spaces are the
product of the Borel sets of the components (a \emph{Polish space} is
a topological space which has a countable base and for which a
complete metric exists). The same applies to analytic spaces (an
\emph{analytic space} is a separable metric space which is the continuous image of a Polish space), since
the topology of these these spaces is also countably
generated.

In summary, the observation on products mentioned above suggests that we have to exercise particular care when working with the product of two measurable spaces, which carry a topological structure as well. 

   Given the measurable spaces $(S, {\cal S})$ and $(T, {\cal T})$, a
   map $f: S \to T$ is said to be
   \emph{${\cal S}$-${\cal T}$-measurable} iff $\InvBild{f}{D}\in{\cal
     S}$ for all $D\in {\cal T}$. Call the measurable map $f: (S,
   {\cal S}) \to (T, {\cal T})$
   \emph{final} iff ${\cal T}$ is the final $\sigma$-algebra with
   respect to $f$ and ${\cal S}$. Hence $f$ is final iff ${\cal T}$
   is the largest $\sigma$-algebra $\mathcal{C}$ on $T$ such that
   $\InvBild{f}{\mathcal{C}} := \{\InvBild{f}{C}\mid C\in {\cal C}\}\subseteq {\cal S}$ holds, so that $
   {\cal T}= \{B \subseteq T \mid \InvBild{f}{B}\in{\cal S}\}. $ Hence
   we may conclude from $\InvBild{f}{B}\in{\cal S}$ that $B \in {\cal
     T}$.  An equivalent formulation for finality of $f$ is that a map
   $g: T \to U$ is ${\cal T}$-${\cal U}$-measurable if and only if
   $g\circ f: S\to U$ is ${\cal S}$-${\cal U}$-measurable, whenever
   $(U, {\cal U})$ is a measurable space. Measurability of real valued
   maps always refers to the Borel sets on the reals, hence $f: S\to
   \Real$ is measurable iff $\{s\in S\mid f(s)\bowtie q\}\in{\cal S}$
   for each rational number $q$, with $\bowtie$ as one of the
   relations $\leq, <, \geq, >$.  

Measurable spaces with measurable maps as morphisms form a
category, which, however, does not have an extra symbol assigned to
it in the present paper.

\medskip

Let $\rho$ be an equivalence relation on $S$. Call $A \subseteq S$
an \emph{$\rho$-invariant set} iff $A$ is the union of $\rho$-classes,
equivalently, iff $s \in A$ and $\isEquiv{s}{s'}{\rho}$ implies
$s'\in A$. 
Then
\begin{equation*}
\Inv{\rho}{{\cal S}} := \{A \in {\cal S} \mid A \text{ is $\rho$-invariant}\}
\end{equation*}
denotes the $\sigma$-algebra of $\rho$-invariant measurable subsets
of $S$.  As usual,
\begin{equation*}
\Kern{f} := \{\langle s, s'\rangle \mid f(s) = f(s')\}  
\end{equation*}
is the \emph{kernel of $f$}.

\medskip

We write $\SubProb{S, {\cal S}}$ for the set of all subprobability
measures on the measurable space $(S, {\cal S})$. This space
is made a measurable space upon taking as a $\sigma$-algebra
\begin{equation}
\label{intro-weak-sigma}
w({\cal S}) := \sigma\bigl(\{\basS{(S, {\mathcal S})}{A}{\bowtie q} \mid A \in
{\cal S}, q \in [0, 1]\}\bigr)
\end{equation}
Here 
\begin{equation*}
 \basS{(S, {\mathcal S})}{A}{\bowtie q} := \{\mu \in \SubProb{S, {\cal S}} \mid \mu(A) \bowtie q\}
\end{equation*}
is the set of all subprobabilities on $(S, {\cal S})$ which evaluate on the measurable set $A$ as
$\bowtie q$, where $\bowtie$ is one of the relations $\leq, <,\geq,
>$. This $\sigma$-algebra is sometimes called the
\emph{weak-*-$\sigma$-algebra}. 

A morphism  $f: (S, {\cal S}) \to (T, {\cal T})$ in the category of
measurable spaces induces a map 
$
\SubProbSenza{f}: \SubProb{S, {\cal S}}\to\SubProb{T,{\cal T}}
$
upon setting 
\begin{equation*}
(\SubProbSenza{f})(\nu)(B) := \nu(\InvBild{f}{B})
\end{equation*}
for $B \in \Borel{T, {\cal T}}$; as usual, $\SubProbSenza{f}$ is
sometimes written as $\SubProb{f}$. Because $
\InvBild{(\SubProbSenza{f})}{\basS{T}{B}{\bowtie q}} =
\basS{S}{\InvBild{f}{B}}{\bowtie q}, $ this map is $w({\cal
  S})$-$w({\cal T})$-measurable as well. Thus $\SubProbSenza$ is an
endofunctor on the category of measurable spaces with measurable maps
as morphisms; in fact, it is the functorial part of a monad which is
sometimes called the Giry monad~\cite{Giry}, for a slight extension
see~\cite{EED-PipesAndFilters}.

\medskip

From now on, we will not write down explicitly the $\sigma$-algebra
${\cal S}$ of a measurable space $(S, {\cal S})$, unless there is good
reason to do so. Furthermore the space $\SubProb{S}$ of all
subprobabilities will be understood to carry the
weak-*-$\sigma$-algebra $w({\cal S})$ \textbf{always}. We will write $\Sigma_{\rho}$ for $\Inv{\rho}{{\cal S}}$, and $\Sigma_{\Kern{f}}$ will be abbreviated as $\Sigma_{f}$.

\BeginDefinition{stoch-relation}
Given two measurable spaces $S$ and $T$, a \emph{stochastic relation} (or
\emph{sub Markov kernel}) $K: S\Trans T$ from $S$ to $T$ is a measurable map $S\to
\SubProb{T}$. 
\EndDefinition

$K: S\Trans T$ is a stochastic relation iff these conditions
hold
\begin{enumerate}
\item $K(s)$ is for each $s\in S$ a subprobability measure on the
  $\sigma$-algebra ${\cal T}$ of $T$.
\item For each $D\in {\cal T}$, the map $s\mapsto K(s)(D)$ is
  measurable. 
\end{enumerate}
This characterization is well known. A stochastic relation $K: S\Trans S$ models probabilistic transitions: $K(s)(C)$ is interpreted as the probability that the next state is a member of $D$ after making a transition from $s$; if $K(s)(S) < 1$, the event that there is no next state may occur with positive probability. 

It can be shown that stochastic relations are the Kleisli morphisms
for the Giry monad~\cite{Giry}.

\subsection{Some Indispensable Tools}
\label{sec:indispensable}

We post here for the reader's convenience some measure theoretic tools which will be used all over. Fix a set $S$. 

\paragraph{Dynkin's $\pi$-$\lambda$-Theorem.}
This technical tool is most useful when it comes to determine the $\sigma$-algebra generated by a family of sets~\cite[Theorem 10.1]{Kechris}. 

\BeginProposition{pi-lambda}
Let $\mathcal{A}$ be a family of subsets of $S$ that is closed under finite intersections. Then $\sigma(\mathcal{A})$ is the smallest family of subsets containing $\mathcal{A}$ which is closed under complementation and countable disjoint unions. 
\QED
\EndProposition

\paragraph{Choquet's Representation.}
The following condition on product measurability and an associated integral representation attributed to Choquet is used~\cite[Corollary 3.4.3]{Bogachev}. Assume that $(S, {\mathcal S})$ is a measurable space. 
\BeginTheorem{Choquet}
Let $f: S \to \pReal$ be measurable and bounded, then
\begin{equation}
\label{Choquet-1}
C_{\bowtie}(f) := \{\langle s, r\rangle \in S \times \pReal \mid f(s) \bowtie r\} \in {\mathcal S}\otimes\Borel{\pReal}.
\end{equation}
If $\mu$ is a $\sigma$-finite measure on ${\mathcal S}$, then
\begin{equation}
\label{Choquet-2}
\int_S f(s)\ \mu(dx)  = \int_0^\infty \mu(\{s \in S \mid f(x) > t\})\ dt
= (\mu\otimes\lambda)(C_{>}(f)).
\end{equation}
with $\mu\otimes\lambda$ as the product of $\mu$ with Lebesgue measure $\lambda$.
\QED
\EndTheorem
For $S$ an interval in $\Real$, the set 
$
C_>(f) = \{\langle s, t\rangle \in S \times \pReal \mid 0 \leq t < f(s)\}
$
may be visualized as the area between the $x$-axis and the graph of $f$. Hence formula~(\ref{Choquet-2}) specializes to the Riemann integral, if $f: \pReal\to\pReal$ is Riemann integrable, and $\mu$ is also Lebesgue measure.
\subsection{Tame Relations}
\label{sec:tame-relations}
A $\sigma$-algebra $\mathcal{A}\subseteq{\mathcal S}$ on the measurable space $(S,{\mathcal S})$ induces an equivalence relation $\rho_{\mathcal{A}}$ upon setting 
\begin{equation}
\label{det-equiv}
\isEquiv{s}{s'}{\rho_{\mathcal{A}}} :\Longleftrightarrow [\forall A \in \mathcal{A}_0: s \in A \text{ iff } s'\in A]
\end{equation}
for some generator $\mathcal{A}_0$ of $\mathcal{A}$ (${\cal A}_0$ may be ${\cal A}$, of course). It is easy to see that each element of $\mathcal{A}$ is $\rho_{\mathcal{A}}$-invariant. But we do not have necessarily 
$
\Sigma_{\rho_{\mathcal{A}}} = \mathcal{A}:
$
Take for example $S$ as the reals $\Real$, where the Borel sets $\Borel{\Real}$ are taken as the $\sigma$-algebra, and take $\mathcal{Y}$ as the countable-cocountable sub-$\sigma$-algebra of $\Borel{\Real}$, then $\rho_{\mathcal{Y}}$ is the identity, and 
$
\Sigma_{\rho_{\mathcal{Y}}} = \Borel{\Real}.
$
\BeginDefinition{exact}
Given an equivalence relation $\rho$ and a subset $\mathcal{A}\subseteq{\mathcal S}$ of the measurable sets of $S$, we call \emph{$\rho$ exact with $\mathcal{A}$} iff 
$
\Sigma_{\rho} = \sigma(\mathcal{A}).
$
\EndDefinition
Thus $\mathcal{A}$ generates \emph{exactly} the invariant measurable sets of $\rho$, if $\rho$ is exact with $\mathcal{A}$; this notion will be helpful below. It is easy to see that $\mathcal{A}$ determines $\rho$ as in~(\ref{det-equiv}). Taking $\mathcal{Y}$ as above, we see that $\rho_{\mathcal{Y}}$ is not exact with $\mathcal{Y}$; it is, however, exact with the open sets $\tau$ or the intervals $\mathcal{I}$ of $\Real$, because 
$
\Sigma_{\rho_{\mathcal{Y}}} = \Borel{\Real} = \sigma(\tau) = \sigma(\mathcal{I}).
$

The set $\Faktor{S}{\rho}$ of all equivalence classes is endowed with the final $\sigma$-algebra $\Faktor{{\mathcal S}}{\rho}$ with respect to the factor map 
$
\fMap{\rho}: s \mapsto \Klasse{s}{\rho},
$
i.e., the largest $\sigma$-algebra rendering $\fMap{\rho}$ measurable. Hence 
\begin{equation*}
\Faktor{{\mathcal S}}{\rho} = \{C \subseteq \Faktor{S}{\rho} \mid \InvBild{\fMap{\rho}}{C} \in{\mathcal S}\}.
\end{equation*}
It follows that $\Bild{\fMap{\rho}}{B}\in\Faktor{{\mathcal S}}{\rho}$ whenever $B \in \Sigma_{\rho}$, because 
$
B = \InvBild{\fMap{\rho}}{\Bild{\fMap{\rho}}{B}}
$
on account of the invariance of $B$. 

The first part of the following statement is obvious.
\BeginLemma{kerf-is-exact}
Let $f: S \to T$ be a measurable map, then $\Kern{f}$ is exact with $\Sigma_{f}$. If $f$ is final, then 
$
\{\InvBild{f}{E}\mid E\in {\mathcal T}\} =\Sigma_{f}.
$
\EndLemma

{
\def\fKf{\fMap{\Kern{f}}}
\def\xKf{\Faktor{S}{\Kern{f}}}
\BeginProof
1.
If $E\in{\mathcal T}$, we know that $\InvBild{f}{E}\in{\mathcal S}$. Since $\InvBild{f}{E}$ is $\Kern{f}$-invariant, we conclude that  
$
\{\InvBild{f}{E}\mid E\in {\mathcal T}\} \subseteq \Sigma_{f}.
$

2.
For the other inclusion, decompose $f$ as $\widetilde{f}\circ \fKf$, then $\widetilde{f}: \xKf\to T$ is injective, and, since the domain carries a final $\sigma$-algebra, it is measurable. Because $\widetilde{f}$ is injective, the inverse $\widetilde{f}^{-1}: \PowerSet{T}\to \PowerSet{\xKf}$ is onto. Now let $D\in\Sigma_{f}$, then $\Bild{\fKf}{D} \in\Faktor{{\mathcal S}}{\Kern{f}}$, because 
$
 D = \InvBild{\fMap{\Kern{f}}}{\Bild{\fMap{\Kern{f}}}{D}},
$
since $D$ is $\Kern{f}$-invariant. Because $\widetilde{f}^{-1}$ is onto, we find some $E\subseteq T$ with
$
\Bild{\fMap{\Kern{f}}}{D} = \InvBild{\widetilde{f}}{E},
$
and we conclude 
$
D = \InvBild{\fMap{\Kern{f}}}{\Bild{\fMap{\Kern{f}}}{D}} = \InvBild{\fMap{\Kern{f}}}{\InvBild{\widetilde{f}}{E}} = \InvBild{f}{E},
$ 
so that $E \in {\mathcal T}$, since in particular  $D\in{\mathcal S}$. This implies
$
\Sigma_{f}\subseteq \{\InvBild{f}{E}\mid E\in{\mathcal T}\}.
$
\EndProof
}

\medskip

Let us briefly mention an important special case. Assume that $S$ and
$T$ are Polish, then each Borel measurable and
surjective map $f: S\to T$ is final~\cite[Proof of Lemma
1.7.10]{EED-CoalgLogic-Book}. Hence final maps occur in a fairly
natural way in a topological setting.

The following observations are helpful consequences. We first show that $\Sigma_{f}$ and ${\cal T}$ have the same structure as Boolean $\sigma$-algebras, if $f:S\to T$ is final and onto.

\BeginCorollary{isomorphism-f-final}
Let $f: S\to T$ be final and surjective. Then $f^{-1}$ is an isomorphism of the $\sigma$-algebras $\Sigma_{f}$  and ${\cal T}$. 
\EndCorollary

\BeginProof
Because $f$ is onto, $f^{-1}: {\cal T}\to \Sigma_{f}$ is
injective. Now let $E\in\Sigma_{f}$; we claim that
$\InvBild{f}{\Bild{f}{E}} = E$. In fact, let $f(s)\in \Bild{f}{E}$,
then there exists $s'\in E$ with $f(s) = f(s')$. Since $E$ is
$\Kern{f}$-invariant, we conclude $s\in E$, hence
$\InvBild{f}{\Bild{f}{E}}\subseteq E$. The other inclusion is
trivial. Because $f$ is final, we conclude from $\InvBild{f}{\Bild{f}{E}}\in \Sigma_{f}\subseteq{\mathcal S}$ that $\Bild{f}{E} \in {\cal
  T}$. Hence $f^{-1}$ is onto as well, so it is a bijection. Because
$f^{-1}$ is compatible with finite or countable Boolean
operations, we have established that $f^{-1}$ is the isomorphism we are
looking for.
\EndProof

We could have used the induced map $f: \Sigma_{f}\to {\mathcal T}$ in
the scenario above and then have shown that $f$ is an isomorphism. The
crucial observation is that $\Bild{f}{\bigcap_{i\in I} A_{i}} =
\bigcap_{i\in I}\Bild{f}{A_{i}}$ holds for any family
$\bigl(A_{i}\bigr)_{i\in I}$ in $\Sigma_{f}$. In fact, assume that
$I\not=\emptyset$, and let $t\in \bigcap_{i\in
  I}\Bild{f}{A_{i}}$. Then there exists for each $j\in I$ an element
$a_{j}\in A_{i}$ with $t=f(a_{j})$. Because all $A_{i}$ are
$\Kern{f}$-invariant, one concludes that $a_{j}\in\bigcap_{i\in
  I}A_{i}$ for all $j\in I$, hence $t\in \Bild{f}{\bigcap_{i\in I}
  A_{i}}$; the other inclusion is trivial. In the same manner one sees
that $\Bild{f}{S\setminus A} = T\setminus\Bild{f}{A}$ for $A\in
\Sigma_{f}$. Hence the direct image can be used for the proof of
Corollary~\ref{isomorphism-f-final} as well. But usually the inverse
image is more convenient to work with whenever measures are concerned.

From Corollary~\ref{isomorphism-f-final} we obtain as a consequence that the measure spaces $\SubProb{S, \Sigma_{f}}$ and $\SubProb{T, {\mathcal T}}$ are isomorphic as measurable spaces. Quite apart of being of independent interest, we will use this observation for investigating subprobabilities on one space through those on the other space.  

\BeginCorollary{isom-meas-space}
Let $f: S\to T$ be final and surjective. Then $\SubProb{f}:
\SubProb{S, \Sigma_{f}}\to \SubProb{T, {\mathcal T}}$ is an isomorphism for the measurable spaces  with the respective weak-*-$\sigma$-algebras.
\EndCorollary

\BeginProof
The inverse image map $f^{-1}: {\mathcal T}\to \Sigma_{f}$ is an isomorphism of the Boolean
$\sigma$-algebras by Corollary~\ref{isomorphism-f-final}. Thus, given
$\nu\in\SubProb{T, {\mathcal T}}$, $\mu(C) := \nu(\Bild{f}{C})$
defines a subprobability measure on $(S, \Sigma_{f})$ with
$\SubProb{f}(\mu) = \nu$, because $\SubProb{f}(\mu)(E) =
\mu(\InvBild{f}{E}) = \nu(\Bild{f}{\InvBild{f}{E}}) = \nu(E)$. Thus $\SubProb{f}$ is surjective; it is injective as well, since $\SubProb{f}(\mu_{1})(E)  \not= \SubProb{f}(\mu_{2})(E)$ means that $\mu_{1}$ and $\mu_{2}$ differ on $\InvBild{f}{E}\in\Sigma_{f}$, hence are different members of $\SubProb{S, \Sigma_{f}}$. The generators of the weak-*-$\sigma$-algebras are in a bijective correspondence with each other, which implies that $\SubProb{S, \Sigma_{f}}$ and $\SubProb{T, {\mathcal T}}$ are also isomorphic as measurable spaces. 
\EndProof




\medskip

If $\rho$ is exact with $\mathcal{A}$, we have a handle on the
elements of $\Faktor{{\mathcal S}}{\rho}$, albeit in a special
situation. The characterization below is very similar to Corollary
2.6.5 in~\cite{EED-CoalgLogic-Book}. That statement deals with validity
sets of the formulas of a negation free logic which is closed under
finite conjunctions. The proof given there carries over easily to the
situation at hand.

\BeginLemma{exact-generate}
Let $\rho$ be exact with $\mathcal{A}$, and assume that $\mathcal{A}$ is closed under finite intersections. Then we have
\begin{enumerate}
\item $\Sigma_{\rho} = \sigma(\mathcal{A})$,
\item $\Faktor{{\mathcal S}}{\rho} = \sigma(\{B \subseteq \Faktor{S}{\rho} \mid \InvBild{\fMap{\rho}}{B} \in \mathcal{A}\}).$
\QED
\end{enumerate}
\EndLemma

We require a slightly stronger condition on the equivalence relations we are dealing with, because we need to consider reals in the unit interval $[0, 1]$ as well. Define for this the equivalence relation $\rho\times\Delta$ on $S \times [0, 1]$ upon setting
\begin{equation*}
\isEquiv{\langle s, q\rangle}{\langle s', q'\rangle}{(\rho\times\Delta)} \text{ iff } \isEquiv{s}{s'}{\rho}\text{ and } q = q'.
\end{equation*}

Let us call an equivalence relation $\rho$ tame if the invariant sets
of $\rho\times\Delta$ behave well. In descriptive set theory,
countably generated equivalence relations on a Polish space are
sometimes called tame (they are called \emph{smooth}
in~\cite{EED-CoalgLogic-Book}); the behavior of the present tame
relations is modeled after them. The tame relations in the present paper are quite a different concept from the one discussed by Jacobs in~\cite{Jacobs-tame}.
 
\BeginDefinition{tame}
Call an equivalence relation $\rho$ on the measurable space $(S, {\mathcal S})$ \emph{tame} iff these conditions hold
\begin{enumerate}
\item $\rho$ is exact with some $\mathcal{A}\subseteq {\mathcal S}$,
\item $\rho\times\Delta$ is exact with $\{A \times I \mid A\in
  \mathcal{A}, I\in \Borel{[0, 1]}\}$
\end{enumerate}
\EndDefinition

Thus tameness of $\rho$ tells us that the invariant sets of $\rho\times\Delta$ can just be generated through $\Sigma_{\rho}$ and $\Borel{[0, 1]}$, the invariant sets for $\Delta$. 
Consequently, dealing with the invariant sets for $\rho\times\Delta$ becomes more practical through tameness (\cite[Lemma~3.8]{EED-GameLogic-TR}). Just for the record:
\BeginLemma{is-tame}
The equivalence relation $\rho$ is tame iff
$
\Sigma_{\rho\times\Delta} = \Sigma_{\rho}\otimes\Borel{[0, 1]}
$
holds.
\QED
\EndLemma


Thus we can characterize the factor space with respect to $\rho\times\Delta$ easily; for a proof see~\cite[Corollary 3.9]{EED-GameLogic-TR}.

\BeginCorollary{cor-is-tame}
Assume that $\rho$ is tame, then 
$
\Faktor{(S\otimes [0, 1])}{\rho\times\Delta}
$
and
$
\Faktor{S}{\rho}\otimes[0, 1]
$
are isomorphic as measurable spaces.
\QED
\EndCorollary

We digress briefly and establish tameness in an interesting special
case. Recall that a \emph{smooth equivalence relation} $\rho$ on $S$
has a countable set ${\cal G}\subseteq{\mathcal S}$ such that $
\Sigma_{\rho}= \sigma({\cal G}).  $ Hence smooth equivalence
relations are countably generated. A concise discussion of the properties of
these relations can be found in~\cite[Section
1.7]{EED-CoalgLogic-Book}.

\BeginProposition{smooth-is-tame}
If $\rho$ is smooth, and $S$ is Polish, then $\rho$ is tame.
\EndProposition

\BeginProof
Since $S$ is Polish and $\rho$ is countably generated, the
factor space $\Faktor{S}{\rho}$ is an analytic space~\cite[Proposition
1.7.5]{EED-CoalgLogic-Book}, so in particular
a Hausdorff topological space with $\Borel{\Faktor{S}{\rho}} := \Faktor{\Borel{S}}{\rho}$ as its Borel sets. We infer from~\cite[Lemma 6.4.2
(i)]{Bogachev} that 
$
\Borel{\Faktor{S}{\rho}\otimes[0, 1]} =
\Borel{\Faktor{S}{\rho}}\otimes\Borel{[0, 1]},
$
because $[0, 1]$ is Polish. It is easy to see that $\Sigma_{\rho}\otimes\Borel{[0, 1]}\subseteq\Sigma_{\rho\times\Delta}$, because each
measurable rectangle $A \times B \in \Sigma_{\rho}\otimes\Borel{[0,
  1]}$ is a $\rho\times\Delta$-invariant measurable set, hence
$A\times B \in \Sigma_{\rho\times\Delta}$. We claim that 
$
\InvBild{\fMap{\rho\times\Delta}}{H}\in\Sigma_{\rho}\otimes\Borel{[0, 1]}
$
for all 
$
H\in \Borel{\Faktor{S}{\rho}}\otimes\Borel{[0, 1]}.
$

In fact, let ${\cal H}$ be the set of all elements of
$\Borel{\Faktor{S}{\rho}}\otimes\Borel{[0, 1]}$ for which this is
true. Then ${\cal H}$ is a $\sigma$-algebra, and if 
$
H = A \times B
$
is a rectangle with 
$
A \in \Borel{\Faktor{S}{\rho}}, B \in \Borel{[0, 1]},
$
then 
\begin{equation*}
  \InvBild{\fMap{\rho\times\Delta}}{H} =
  \InvBild{\fMap{\rho}}{A}\times B \in \Sigma_{\rho}\otimes\Borel{[0, 1]}.
\end{equation*}
Thus ${\cal H}$ contains all measurable rectangles which generate the
product $\sigma$-algebra, hence ${\cal H}$ equals $\Borel{\Faktor{S}{\rho}}\otimes\Borel{[0,
  1]}$. 
Now let 
$
D \in \Sigma_{\rho\times\Delta},
$
then
\begin{equation*}
\Bild{\fMap{\rho\times\Delta}}{D}\in\Borel{\Faktor{S}{\rho}}\otimes\Borel{[0,
  1]} = \Borel{\Faktor{S}{\rho}\otimes[0, 1]},
\end{equation*}
hence
\begin{equation*}
D = \InvBild{\fMap{\rho\times\Delta}}{\Bild{\fMap{\rho\times\Delta}}{D}} \in
\Sigma_{\rho}\otimes\Borel{[0, 1]}.
\end{equation*}
This establishes the other inclusion. 
\EndProof

This shows that tame equivalence relations constitute a generalization of smooth ones for the case that we do not work in a Polish environment. Smooth relations help in establishing interesting structural properties for stochastic relations, so it is to be expected that tame relations help in uncovering structural properties for the case of stochastic effectivity functions, to be discussed below. Specifically, we will deal with tame relations when we investigate congruences for
stochastic effectivity functions. 

\section{Stochastic Effectivity Functions}
\label{sec:stoch-effect-funct}

The basic idea for an effectivity function is to produce upon some
input $s$ all the results which can be achieved through $s$. The
results are modelled as subsets of some result set, so one of the
basic requirements is that the family of sets thus achieved is upward closed:
If $A$ is a set which can be achieved, and $A \subseteq A'$, then it
should be possible to achieve $A'$ as well. Since we are in the realm of probabilities, we do not work
directly with possible outcomes but rather with their distributions. So an
effectivity function should produce an upward
closed set of distributions over the space of outputs upon an input. 

Associating the use of effectivity function with an action, the
sequential composition of actions becomes important. Hence we want to
be able to characterize the results achieved after the execution of
two actions in sequence. This requires taking intermediate results
into account: executing action $\gamma$ will achieve certain results
which then will be fed into the effectivity function associated with
action $\delta$, yielding then the overall result for executing
$\gamma;\delta$.  Again, because we are working in a probabilistic
scenario, we will want to be able to average over intermediate
results. This in turn requires a notion of measurability, permitting
quantitative assessments. 

These considerations, which have been formulated in the Introduction
as requirements, will now be made specific. They lead to the
introduction of stochastic effectivity functions over two measurable
spaces. The latter can be interpreted as input resp. as output space,
but when applying effectivity functions to game logic, the spaces
coincide, forming the state space of the model under consideration. In
order to be better able to distinguish the r\^oles of the domain and
the range spaces, however, we separate both spaces for the purposes of
the present paper.

Effectivity functions are also compared to non-deterministic Markov
processes. We state elementary properties and investigate the
relationship to stochastic relations, i.e., to the transition
probabilities which are at the center of stochastic processes in
probability theory, and which form the basis for stochastic Kripke
models~\cite{Desharnais-Edalat-Panangaden, EED-Book, Panangaden-book}.

Put for the measurable space $(T, {\mathcal T})$ 
\begin{equation*}
  \Vau{T} := \{V \subseteq w({\mathcal T})\mid V \text{ is upward closed}\}
\end{equation*} 
thus if $V \in \Vau{T}$, then $V$ is a collection of  measurable sets of subprobabilities on the measurable space $(T, {\mathcal T})$, moreover $A \in V$ and $A \subseteq B$ together
imply $B\in V$ (for the definition of $w({\mathcal T})$ see~(\ref{intro-weak-sigma})). A measurable map $f: S \to T$ induces a map $\Vau{f}:
\Vau{S}\to\Vau{T}$ upon setting
\begin{equation*}
(\VauSenza{f})(V) := \Vau{f}(V) := \{W \in w({\mathcal T}) \mid \InvBild{\SubProb{f}}{W}\in V\}
\end{equation*}
for $V \in \Vau{S}$, hence clearly $(\VauSenza{f})(V)\in\Vau{T}$. 

Note that $\Vau{T}$ has not been equipped with a $\sigma$-algebra, so
the usual notion of measurability between measurable spaces cannot be
applied; this is in contrast to the non-deterministic labeled Markov
processes studied in~\cite{Terraf-Bisim-MSCS}. 

To be specific,~\cite[Definition~3.1]{Terraf-Bisim-MSCS} defines for a
given measurable space $(S, {\mathcal S})$ of states and a set $L$ of
labels a \emph{non-deterministic labeled Markov process (NLMP)} as a
family $(\kappa_a)_{a\in L}$ of measurable maps $\kappa_a: S\to
w({\mathcal S})$. The target space $w({\mathcal S})$ is endowed with
the smallest $\sigma$-algebra $\mathcal{R}_S$ which contains the sets
$ \{H_G \mid G \in w({\mathcal S})\}, $ where
\begin{equation*}
H_G := \{C \in w({\mathcal S}) \mid C \cap G \not= \emptyset\}
\end{equation*}
is the set of all measurable sets of $\SubProb{S}$ which \emph{hit} the given
Borel set $G \subseteq \SubProb{S}$ (consequently, this $\sigma$-algebra is called the \emph{hit-$\sigma$-algebra}). So, for example $ C \in
H_{\basS{S}{A}{> q}} $ iff $C$ contains some $\mu\in\SubProb{S}$ with
$\mu(A)> q$. It is easy to see that the $\sigma$-algebra $\mathcal{R}_S$
is generated also by the upward closed sets $ U_G := \{C \in
w({\mathcal S})\mid G \subseteq C\} $ for $G \in
w({\mathcal S})$, so that $\kappa_a: S \to w({\mathcal S})$ is
measurable iff $ \InvBild{{\kappa_a}}{U_G}\in\Borel{S} $ holds, whenever
$G \in w({\mathcal S})$. The interplay between NLMPs and stochastic effectivity functions is investigated in~\cite{EED+PST}. 

Concerning $\VauSenza$, a closer look shows that $\Vau{T}$ can be derived from the composition $(\WeSenza\circ \SubProbSenza)(T)$ with
\begin{equation}
\label{label:def_w-set}
  \We{X} := \{W\subseteq\PowerSet{X}\mid W\text{ is upward closed}\}
\end{equation}
for the set $X$, restricting $\WeSenza$ to upward closed subsets of $\sigma$-algebras. We define for the map $f: X\to Y$
\begin{equation}
  \label{label:def_w-map}
  \We{f}(W) := \{Z\in\PowerSet{Y}\mid \InvBild{f}{Z}\in W\}.
\end{equation}
Put for $x\in X$, $F: X\to \We{Y}$
\begin{align*}
  \xi_{X}(x) & := \{W\subseteq X \mid x\in W\},\\
F^{*}(W) & := \bigl\{B\subseteq Y\mid \{x\in X \mid B\in F(x)\}\in W\bigr\},
\end{align*}
then it is not difficult to show~\cite[Example 1.53]{EED-Categs} that $(\WeSenza, -^{*}, \xi)$ is a Kleisli tripel~\cite{Moggi-Inf+Control}, hence forms a monad. 

Let $H \in \bSS{w({\mathcal T})}$ be a measurable subset of $\SubProb{T}\times[0, 1]$ indicating a
\emph{quantitative assessment of subprobabilities}. A typical example could
be 
\begin{equation*}
\{\langle \mu, q\rangle \mid \mu \in \basS{T}{A}{\geq q},\ 0 \leq q \leq 1\} 
= 
\{\langle \mu, q\rangle \mid \mu(A) \geq q, \ 0 \leq q \leq 1\}
\end{equation*}
for some $A \in {\mathcal T}$, asking for all combinations of
subprobabilities and reals such that the probability for the given set
$A$ of states or events do not lie below this value. Consider a map $P: S \to \Vau{T}$, fix some real $q$ and consider the
\emph{horizontal section} $H_q = \{\mu \mid \langle \mu, q\rangle \in H\}$ of $H$ at $q$, viz., the set 
of all measures evaluated through $q$. We ask for all states $s$ such that $H_{q}$ is \emph{effective for $s$}, i.e., 
$
\{s \in S \mid H_q \in P(s)\}. 
$
This set should be a measurable subset of $S$; an NLMP will have this property after a simple transformation, see below and~\cite{EED+PST}. It turns out, however, that this is not enough, we also require the real components being captured through a measurable set as well --- after all, the real component will be used to be averaged over later on, so it should behave decently. This idea is captured in the following definition.

\BeginDefinition{t-measurability}
Call a map $P: S \to \Vau{T}$ \emph{t-measurable} iff 
$
\{\langle s, q\rangle \mid H_q\in P(s)\} \in \bSS{{\mathcal S}}\}
$
whenever $ H \in\bSS{w({\mathcal T})} $ 
\EndDefinition

Thus if $P$ is t-measurable then we know in particular that all pairs of states and
numerical values indicating the effectivity of the evaluation of a
measurable set $A\in {\mathcal T}$, i.e., the set $ \{\langle s,
q\rangle \mid \basS{T}{A}{\bowtie q} \in P(s)\} $ is always a
measurable subset of $S\otimes[0, 1]$. This is so because we know that
\begin{equation*}
\{\langle \mu, q \rangle \mid \mu\in \basS{T}{A}{\bowtie q}, 0 \leq q
\leq 1\}
=
\{\langle \mu, q\rangle \in \SubProb{T}\times[0, 1]\mid \mu(A) \bowtie q\} \in \bSS{w({\mathcal T})}
\end{equation*}
by Theorem~\ref{Choquet}. 
\medskip

This leads to the notion of a stochastic effectivity function.

\BeginDefinition{effFnct} 
Given measurable spaces $S$ and $T$, a \emph{stochastic effectivity
  function} $P: S \eTrans T$ from $S$ to $T$ is a t-measurable map $P:
S \to\Vau{T}$.
\EndDefinition

We say that, given $P: S \eTrans T$, the set $P(s)$ comprises the
\emph{portfolio} for $s\in S$, and that each element of $P(s)$ can be
\emph{achieved} or is \emph{effective through $P$ in $s$}; sometimes $S$ may be
thought to be a set of inputs, and  the portfolio to be defined over the set of outputs which can be achieved. If $S$ and $T$ coincide, representing a set of states, $P(s)$ may be interpreted as indicating the set of all state distributions which can be achieved in state $s$. 
  
The kinship to NLMPs is fairly obvious. Let $P: S \eTrans S$ be a stochastic effectivity function on a state
space $S$, then $P(s)$ is an upward closed \emph{subset} of
$w({\mathcal S})$ for any $s \in S$. If
$(\kappa_a)_{a\in L}$ is a non-deterministic labeled Markov process on
the same state space, then $\kappa_a(s)$ is an \emph{element} of
$w({\mathcal S})$. It can be made a map 
$
\kappa^*_a: S \to \Vau{S}
$
upon setting 
\begin{equation*}
  \kappa^*_a(s) := \{A \in w({\mathcal S}) \mid \kappa_a(s)\subseteq A\}.
\end{equation*}
It is not difficult to see that the set $ \{s \in S \mid H \in
\kappa^*_a(s)\} $ is a member of ${\mathcal S}$ whenever $H \in
w({\mathcal S})$, since $\kappa_a: S \to w({\mathcal S})$ is ${\mathcal
  S}$-$\mathcal{R}_S$-measurable. But this does not suffice for our
purposes, since a quantitative assessment is missing. Hence the
kinship is remote only. These concepts are obviously similar in spirit
and intention, viz., to capture stochastic non-determinism.

Let us have a look at some examples.

\BeginExample{finite-cases}
We show that a finite transition system can be converted into a stochastic effectivity function.

Let $S := \{1, \dots, n\}$ for some $n\in\Nat$, and take the power set as a $\sigma$-algebra. Then $\SubProb{S}$ can be identified with the compact convex set
\begin{equation*}
  \Pi_{n} := \{\langle x_{1}, \dots, x_{n}\rangle\mid x_{i}\geq 0 \text{ for }1\leq i \leq n, \sum_{i=1}^{n} x_{i}\leq 1\}.
\end{equation*}
Geometrically, $\Pi_{n}$ is the convex hull of the unit vectors $e_{i}$, $1\leq i
\leq n$ and the zero vector; here $e_{i}(i) = 1$, and $e_{i}(j) = 0$
if $i\not= j$ is the $i$-th $n$-dimensional unit vector. The weak-*-$\sigma$-algebra is the
Borel-$\sigma$-algebra $\Borel{\Pi_{n}}$ for the Euclidean topology on
$\Pi_{n}$. 

Assume we have a transition system $\to_{S}$ on $S$, hence
a relation $\to_{S}\subseteq S\times S$. Put $succ(s) := \{s'\in S\mid
s\to_{S} s'\}$ as the set of a successor states for state $s$. Define for
$s\in S$ the set of weighted successors
\begin{equation*}
  \kappa(s):= \{\sum_{s'\in succ(s)}\alpha_{s'}\cdot e_{s'}\mid \Rational\ni\alpha_{s'}\geq 0 \text{ for } s'\in succ(s), \sum_{s'\in succ(s)}\alpha_{s'}\leq 1\}
\end{equation*}
and the upward closed set
\begin{equation*}
P(s)  := \{A\in\Borel{\Pi_{n}}\mid \kappa(s)\subseteq A\}
\end{equation*}
A set $A$ is in the portfolio for $P$ in state $s$ if $A$  contains all
rational distributions on the successor states. We will restrict our
attention to these rational distributions, which are positive convex
combinations of the unit vectors with rational coefficients. Note that states may get
lost, since we work with subprobabilities.

We claim that $P$ is an effectivity function on $S$. If $P(s)=\emptyset$, there is nothing to show, so we assume that always $P(s)\not=\emptyset$. Let
$H\in\bSS{\Borel{\Pi_{n}}} = \Borel{\Pi_{n}\otimes[0, 1]}$, then
\begin{equation*}
  \{\langle s, q\rangle \mid H_{q}\in P(s)\} = \bigcup_{1\leq s \leq n}\{s\}\times \{q\in[0, 1]\mid H_{q}\in P(s)\}.
\end{equation*}
Fix $s\in S$, and let $succ(s) = \{s_{1}, \dots, s_{m}\}.$ Put
\begin{equation}
\label{def-mn}
\Omega_m := \{\langle \alpha_{1}, \dots, \alpha_{m}\rangle\in\Rational^{m}\mid  \alpha_{i}\geq 0, \sum_{i}\alpha_{i}\leq 1\},
\end{equation}
hence $\Omega_{m}$ is countable, and 
\begin{align*}
  \{q\in[0, 1]\mid H_{q}\in P(s)\}
& = 
\{q\in[0, 1] \mid k(s)\subseteq H_{q}\}\\
& = \bigcap_{\langle \alpha_{1}, \dots, \alpha_{m}\rangle\in \Omega_{m}}\{q\in[0, 1]\mid \sum_{i}\alpha_{i}\cdot e_{j_{i}}\in H_{q}\}.
\end{align*}
Now fix $\alpha := \langle \alpha_{1}, \dots, \alpha_{m}\rangle\in
\Omega_{m}$. The map $\zeta_{\alpha}: [0, 1]^{m\cdot n}\to [0, 1]^{n}$ which maps
$\langle v_{1}, \dots, v_{m}\rangle$ to $\sum_{i=1}^{m}\alpha_{i}\cdot
v_{i}$ is continuous, hence measurable, so is $\xi := \zeta_{\alpha}\times
id_{[0, 1]}: [0, 1]^{m\cdot n}\times[0, 1]\to [0, 1]^{n}\times[0,
1]$. Hence $I := \InvBild{\xi}{H}\in\Borel{[0, 1]^{m\cdot n}\times[0,
  1]}$, and $ \sum_{i=1}^{m}\alpha_{i}\cdot e_{j_{i}}\in H_{q} $ iff $
\langle e_{j_{1}}, \dots, e_{j_{m}}, q\rangle\in I.  $ Consequently, 
\begin{equation*}
\{q\in[0, 1]\mid \sum_{i}\alpha_{i}\cdot e_{j_{i}}\in H_{q}\} =
I^{\langle e_{j_{1}}, \dots, e_{j_{m}}\rangle}\in \Borel{[0, 1]}.
\end{equation*}
But this implies that 
\begin{equation*}
\{q\in[0, 1]\mid H_{q}\in P(s)\} =
\bigcap_{\alpha\in \Omega_{m}}\{q\in[0, 1]\mid \sum_{i}\alpha_{i}\cdot
e_{j_{i}}\in H_{q}\}\in\Borel{[0, 1]}
\end{equation*}
for the fixed state $s\in
S$. Collecting states, we obtain
\begin{equation*}
  \{\langle s, q\rangle\in S\times[0, 1]\mid H_{q}\in P(s)\} \in \PowerSet{S}\otimes\Borel{[0, 1]}.
\end{equation*}
Thus we have converted a finite transition system 
into a stochastic effectivity function by constructing all
subprobabilities over the respective successor sets, albeit with
rational coefficients. It is fairly easy to see that $\kappa$ forms an NLMP. 

One might ask whether the restriction to
rational coefficients is really necessary. Taking the convex
closure with real coefficients might, however, results in loosing measurability,
see~\cite[p. 216]{Kechris}.
\EndExample

The next example shows that a stochastic effectivity function can be used for interpreting a simple modal logic. 

\BeginExample{interpr-simple-modal}
Let $\Phi$ be a set of atomic propositions, and define the formulas of a logic through this grammar
\begin{equation*}
  \phi ::= \top\mid p\mid \phi_{1}\wedge\phi_{2}\mid \Diamond_{q}\phi
\end{equation*}
with $p\in\Phi$ an atomic proposition and $q\in[0, 1]$ a threshold value. Intuitively, $\Diamond_{q}\phi$ is true in a state $s$ iff there can be a move in $s$ to a state in which $\phi$ holds with probability not smaller than $q$. 

This logic is interpreted over a measurable space $(S,\mathcal{S})$;
assume that we are given a map $e: \Phi\to \mathcal{S}$, assigning
each atomic proposition a measurable set as its validity set. Let $P:
S\eTrans S$ be a stochastic effectivity function over $(S,
\mathcal{S})$, then define inductively 
{
 \def\Gilt[#1]{\LinkeKlammer#1\RechteKlammer}
\begin{align*}
  \Gilt[\top] & := S,\\
\Gilt[p] & := e(p),\text{ for }p\in\Phi,\\
\Gilt[\phi_{1}\wedge\phi_{2}] & := \Gilt[\phi_{1}]\cap\Gilt[\phi_{2}],\\
\Gilt[\Diamond_{q}\phi] & := \{s\in S \mid \basS{S}{\Gilt[\phi]}{>q}\in P(s)\}.
\end{align*}

The interesting line is of course the last one. It assigns to
$\Diamond_{q}\phi$ all states $s$ such that $\basS{S}{\Gilt[\phi]}{>q}$ is
in the portfolio of $P(s)$. These are all states for which the collection of all measures yielding an evaluation
on $\Gilt[\phi]$ greater than $q$ can be achieved.

Then t-measurability of $P$ and the
assumption on $e$ make sure that these sets are measurable. This is
shown by induction on the structure of the formulas.  
}
\EndExample

We have a look now at stochastic relations as a concept which
specializes both NLMPs and stochastic effectivity functions.

\BeginExample{ex-stochRel}
Let $K: S \Trans T$ be a stochastic relation, then 
\begin{equation*}
P_K(s) := \{A \in \bost \mid K(s)\in A\}
\end{equation*}
defines a stochastic effectivity function $P_K: S \eTrans
T$~\cite[Lemma 4.4]{EED-GameLogic-TR}. Similarly, assume 
$
{\cal F} = \{K_n \mid n \in \Nat\}
$
is a countable family of stochastic relations $K_n: S \Trans T$, then 
\begin{align*}
  (\exists {\cal F})(s) & := \{A \in \bost \mid K_n(s)\in A\text{ for
    some } n \in \Nat\},\\
 (\forall {\cal F})(s) & := \{A \in \bost \mid K_n(s)\in A\text{ for
    all } n \in \Nat\}
\end{align*}
define stochastic effectivity functions $S \eTrans T$. These functions
resemble the weak resp. the strong inverse of set-valued maps studied
in topology or economics~\cite{Michael,Himmelberg}. 
\EndExample

A pattern seems to arise here: take a hit-measurable map $\kappa: S\to w({\mathcal T})$, and define 
\begin{equation*}
  P_{\kappa}(s) := \{A\in w({\mathcal T}) \mid \kappa(s)\subseteq A\}.
\end{equation*}
Then $P_{\kappa}(s)$ is evidently upward closed, and the example above
suggests that this yields a stochastic effectivity function. It turns
out, however, that this picture has to be scrutinized carefully. If
$\kappa(s)$ is always finite, then $P_{\kappa}$ is a stochastic
effectivity function indeed, i.e., $P_{\kappa}$ is t-measurable. If,
however, $\kappa(s)$ is uncountable for some $s$, t-measurability is
gone forever. This is discussed in some detail in~\cite{EED+PST}.

\BeginExample{ex-stochRel-1} 
Continuing Example~\ref{ex-stochRel}, let $\{1, \dots, \ell\}$ be a
set of individuals, and $K_1, \dots, K_\ell$ be a finite set of
stochastic relations.  $K_j: S \Trans T$ is intended to model the
preferences of individual $j$.
Then the set of all rational positive convex combinations  ($\Omega_{\ell}$ as in~(\ref{def-mn}))
\begin{equation*}
\mathcal{F} := \bigl\{\sum_{j=1}^\ell \alpha_j\cdot K_j\mid \langle \alpha_1, \dots, \alpha_\ell\rangle \in \Omega_{\ell}\bigr\}
\end{equation*}
defines a countable family of stochastic relations $S\Trans T$. The
relation $ \sum_{j=1}^\ell \alpha_j\cdot K_j\ $ indicates a possible
group preference, individual $j$ being assigned weight $\alpha_j$ (the
weights do not necessarily add up to $1$, indicating some possible
loss along the process of coordination).  If we have $ A \in (\exists
{\cal F})(s), $ then the set $A$ of distributions indicates that
portfolio $A$ is possible in state $s$ for certain preferences
$\alpha\in \Omega_{\ell}$, if however $ A \in (\forall {\cal F})(s), $
portfolio $A$ is always preferred. This idea can be extended easily to
a countable number of individuals with finite coalitions by taking a
sequence $\Folge{K}$ of stochastic relations and considering
\begin{equation*}
\Omega' := \{ \Folge{\alpha}\mid \sum_n \alpha_n\leq 1, 0 \leq \alpha_n \text{
    rational}, \alpha_n \not= 0 \text{ for finitely many } n\}
\end{equation*}
rather than $\Omega_{\ell}$. The same \emph{caveat} as in
Example~\ref{finite-cases} applies when it comes to taking the
``full'' convex closure with real coefficients.
\EndExample

\medskip

Effectivity functions which are generated through stochastic relations
can be characterized in terms of principal ultrafilters, as the next
proposition shows.

\BeginProposition{pointed}
Let $P: S \eTrans T$ be a stochastic effectivity function. Then these statements are equivalent
\begin{enumerate}
\item\label{item:1} $P(s)$ is a principal ultrafilter on $w({\mathcal T})$ for each $s \in S$.
\item \label{item:2} $P = P_K$ for some stochastic relation $K: S \Trans T$.
\end{enumerate}
\EndProposition
\BeginProof
We shown only \labelImpl{item:1}{item:2}: Let $P(s)$ be a principal ultrafilter for each $s\in S$, and define
\begin{equation*}
  \{K(s)\} := \bigcap P(s).
\end{equation*}
Then $K(s)\in\SubProb{T}$ for each $s \in S$, and we show now that 
$
\{s \in S \mid K(s)(G) > q\} \in {\mathcal S}
$
for each $G \in {\mathcal T}, 0 \leq q \leq 1$. We claim first that 
$
K(s) \in \basS{T}{G}{> q}
$
is equivalent to 
$
\basS{T}{G}{> q}\in P(s).
$
For assume that $K(s)\in \basS{T}{G}{> q}$ but $\basS{T}{G}{> q}\notin P(s)$. Since $P(s)$ is an ultrafilter,
$
\SubProb{T}\setminus \basS{T}{G}{> q} = \basS{T}{G}{\leq q} \in P(s).
$
But this means 
$
K(s)(G) \leq q,
$
contradicting the assumption. Now we infer
\begin{equation*}
  \{\langle s, q\rangle \mid K(s)(G) > q\}
=
\{\langle s, q\rangle \mid \basS{T}{G}{> q}\in P(s)\} \in\bSS{{\mathcal S}},
\end{equation*}
which implies that we have in particular
$
\{s \in S \mid K(s)(G) > q\} \in {\mathcal S}
$
for fixed $q\in [0, 1]$. 
\EndProof

\medskip

The relationship of stochastic effectivity functions and stochastic relations can be characterized also through an approach resembling deduction systems~\cite{Goldblatt-Deduction, EED-GameLogic-TR}. This is sketched here, the reader is referred to~\cite[Section~4.1]{EED-GameLogic-TR} for details and proofs.

A \emph{characteristic relation} $R \subseteq [0, 1]\times{\mathcal T}$ on the measurable space $(T, {\mathcal T})$ is defined through these conditions
\begin{align*}
\text{ }\ &\frac{\langle r, A\rangle\in R, A \subseteq B}{\langle r, B\rangle\in R} &
\text{ }\  &\frac{\langle r, A\rangle\in R, r \geq s}{\langle s, A\rangle\in R}\\
\text{ }\  &\frac{\langle r, A\rangle\notin R, \langle s, B\rangle\notin R, r + s \leq 1}{\langle r+s, A\cup B\rangle\notin R}&
\text{ }\   &\frac{\langle r, A\cup B\rangle\in R, \langle s,A \cup (S\setminus B)\rangle\in R, r + s \leq 1}{\langle r+s, A\rangle\in R}\\
\text{ }\   &\frac{\langle r, A\rangle\in R, r + s > 1}{\langle s, S \setminus A\rangle\notin R}&
\text{ }\   &\frac{\langle r, \emptyset\rangle\in R}{r = 0}\\
\text{ }\  &\frac{A_1 \supseteq A_2 \supseteq \dots, \forall n \in \Nat: \langle r, A_n\rangle\in R}{\langle r, \bigcap_{n\geq 1} A_n\rangle \in R}
\end{align*}
Such a characteristic relation $R$ defines a subprobability $\mu_R\in \SubProb{T}$ through
\begin{equation*}
\mu_R(B) := \sup\{r \in [0, 1] \mid \langle r, B\rangle\in R\}.
\end{equation*}
An upper closed set $Q \in \Vau{T}$ is said to \emph{satisfy} relation $R$ iff we have
\begin{equation*}
\langle q, A\rangle \in R 
\Leftrightarrow
\basS{T}{A}{\geq q}\in Q,
\end{equation*}
and $Q$ is said to \emph{implement} $\mu \in \SubProb{T}$ iff 
\begin{equation*}
\mu(A) \geq q
\Leftrightarrow
\basS{T}{A}{> q}\in Q.
\end{equation*}
Then $Q$ satisfies the characteristic relation $R$ iff it implements $\mu_R$. Moreover, for a given effectivity function $P: S \eTrans T$ there exists a stochastic relation $K: S \Trans T$ with $P = P_K$ iff 
\begin{equation*}
R(s) := \{\langle r, B\rangle \mid \basS{T}{B}{\geq r} \in P(s)\}
\end{equation*}
defines for each $s \in S$ a characteristic relation such that $R(s)$ satisfies $P(s)$. 

\subsection{Morphisms}
\label{sec:morphisms}

Stochastic effectivity functions $P$ and $Q$ can be compared through
morphisms, which are based on measurable functions. Roughly speaking,
a set is in the portfolio of $Q(f(s))$ iff its inverse image can be
achieved in $P(s)$. This idea is made precise. Having
formulated what morphisms are, we compare them with
morphisms for stochastic relations. Congruences are also defined in
terms of morphisms, see Section~\ref{sec:congruences}.

\medskip

Fix the measurable spaces $(S, {\mathcal S})$, $(T, {\mathcal T})$, $(U, {\mathcal U})$ and $(V, {\mathcal V})$ for the rest of the present paper. 

\medskip

Given stochastic effectivity functions $P: S \eTrans T$ and $Q: U \eTrans V$, a pair of measurable maps $f: S \to U$ and $g: T \to V$ is called a \emph{morphism of effectivity functions} $(f, g): P \to Q$ iff this diagram of maps commutes
\begin{equation*}
\xymatrix{
S\ar[d]_{P}\ar[rr]^f && U\ar[d]^{Q}\\
\Vau{T}\ar[rr]_{\VauSenza{g}} && \Vau{V}
}
\end{equation*}
Thus we have 
\begin{equation}
\label{morph-eff}
W \in Q(f(s))
\Leftrightarrow
\InvBild{(\SubProbSenza{g})}{W}\in P(s)
\end{equation}
for all states $s \in S$ and for all $W \in w({\mathcal U})$; hence
the set $W$ of distributions is in the portfolio of $Q(f(s))$ iff its
inverse image is in the portfolio of $P(s)$. Technically, this
definition derives from the one for functor $\WeSenza$,
see~(\ref{label:def_w-map}).

Let us compare this to morphisms for stochastic relations. A pair of measurable maps $f: S \to U$ and $g: T \to V$ is a \emph{morphism of stochastic relations} $(f, g): K \to L$ for the stochastic relations $K: S \Trans T$ and $L:U \Trans V$ iff this diagram commutes
\begin{equation*}
\xymatrix{
S\ar[d]_{K}\ar[rr]^f && U\ar[d]^{L}\\
\SubProb{T}\ar[rr]_{\SubProbSenza{g}} && \SubProb{V}
}
\end{equation*}
Thus $L(f(s)) = (\SubProbSenza{g})(K(s))$, which means 
\begin{equation}
\label{morph-stoch}
L(f(s))(B) = \bigl(\SubProbSenza{g}\bigr)(K(s))(B) \bigl(= K(s)(\InvBild{g}{B})\bigr)
\end{equation}
for each state $s\in S$ and each measurable set $B \in {\mathcal V}$. This
says that the $L$-distributions of states for $f(s)$ is just the
$K$-distribution of states for $s$, transformed by $\SubProbSenza g$.

These notions of morphisms are related to each other: Each morphism for
stochastic relations turns into a morphism for the associated
effectivity function (we will usually do without the attributions to
effectivity functions or stochastic relations when talking about
morphisms, whenever the context is clear). 
 
\BeginProposition{KripkeToGame}
A morphism $(f, g): K \to L$ for stochastic relations $K$ and $L$ induces a morphism $(f, g): P_K\to P_L$ for the associated stochastic effectivity functions.
\EndProposition

\BeginProof
Fix  a state $s\in S$. Then $W \in P_L(f(s))$ iff $L(f(s))\in W$. Because $(f, g)$ is a morphism $K \to L$, this is equivalent to $(\SubProbSenza{g})(K(s))\in W$, hence to $K(s)\in\InvBild{(\SubProbSenza{g})}{W}$, thus 
$
\InvBild{(\SubProbSenza{g})}{W} \in P_K(s).
$
\EndProof

This result suggests that stochastic effectivity functions are an
algebraically meaningful generalization of stochastic relations.

\subsection{Convolutions}
\label{sec:convolutions}

Given a stochastic effectivity function $P: S \eTrans T$, we ask for
all states $s\in S$ such that $P(s)$ contains a given Borel subset $D$
with a probability not smaller than a threshold value $q$. When we bind
$P$ to an action $\gamma$, this question corresponds to asking for
all states that permit to observe a given effect not below  probability $q$ (e.g., in a game
logic, one might ask for all states such that a formula holds with at
least probability $q$ upon playing a specified game in that state). This
leads to a map ${\mathcal T}\times [0, 1]\to{\mathcal S}$. 

This map is
investigated, and we show that it is possible to model the sequential
composition of actions through this construction; here the assumption
on t-measurability pays off. We give a quick comparison to the
composition of stochastic relations through the Kleisli product and
show how convolution is related to it.
 
Let $P: S \eTrans T$ be a stochastic effectivity function, then 
\begin{equation*}
  \hut{P}(D, q) := \{s \in S \mid \basS{T}{D}{> q} \in P(s)\}
\end{equation*}
defines a  map $\hut{P}$ on ${\mathcal T}\times [0, 1]$ into the power set of $S$. Since $P(s)$ is upward closed for each $s \in S$, the map $\hut{P}(\cdot, q)$ is increasing for each $q$. Now let $D \in {\mathcal T}$, then
\begin{equation*}
 H  := \{ \langle \mu, r \rangle \mid \mu \in  \basS{T}{D}{> r}\} 
 =  \{\langle \mu, r \rangle \mid \mu(D) > r\}
\end{equation*}
is a member of $\bSS{w({\mathcal T})}$ by Choquet's Theorem, thus
\begin{equation*}
  \hut{P}(D, q) = \bigl(\{\langle s, r\rangle \mid H_r \in P(s)\}\bigr)_q \in{\mathcal S}.
\end{equation*}
Hence $\hut{P}$ maps $\mathcal{T}\times [0, 1]$ to ${\mathcal S}$. 

Intuitively, $\hut{P}(D, q)$ gives all states from which it is
possible to achieve a portfolio exceeding the given 
threshold $q$. Binding an action $\gamma$ to $P$ in the sense that $P(s)$ is the
set of all distributions over $T$ upon action $\gamma$ which can be
achieved in state $s$, we interpret $\hut{P}(D, q)$ as the set of all
states $s$ for which we can achieve a state in $D\in\mathcal{T}$ with a probability
greater than $q$. 

Let $Q: T \eTrans U$ be another stochastic effectivity function with
associated map $\hut{Q}: \mathcal{U}\times [0, 1] \to\mathcal{T}.$ Define
the \emph{convolution of $\hut{P}$ and $\hut{Q}$} through
\begin{equation*}
  (\hut{P}\ast\hut{Q})(E, q) := \{s \in S \mid G_Q(E, q) \in P(s)\},
\end{equation*}
where
\begin{equation*}
  G_Q(E, q) := \{\nu \in \SubProb{T} \mid \int_0^1 \nu(\hut{Q}(E, r))\
  dr > q\}.
\end{equation*}
Binding action $\gamma$ to $P$ and $\delta$ to $Q$, respectively, the effect of
executing the combined sequential action $\gamma;\delta$, i.e., first $\gamma$ and then $\delta$, is
modeled through $\hut{P}\ast\hut{Q}$.

Formally, $\hut{P}\ast\hut{Q}$ yields a map from
${\mathcal U}\times[0, 1]$ to ${\mathcal S}$ which is monotone in its first component:

\BeginProposition{conv-measb}
Let $P: S \eTrans T$ and $Q: T \eTrans U$ be stochastic effectivity
functions, then $\hut{P}\ast\hut{Q}$ maps ${\mathcal U}\times[0, 1]$
to ${\mathcal S}$ such that $E
\mapsto\bigl(\hut{P}\ast\hut{Q}\bigr)(E, q)$ is increasing for all
$q$.
\EndProposition

\BeginProof
1.
We show first that 
\begin{equation*}
  G'_Q(E) := \{\langle \nu, q\rangle \mid \int_0^1 \nu(\hut{Q}(E, r))\
  dr > q\} \in \bSS{w({\mathcal T})},
\end{equation*}
whenever $E \in{\mathcal U}$. In order to apply Choquet's
Theorem~\ref{Choquet} for showing that this set is measurable, we
write
\begin{equation*}
  G'_Q(E) = \{\langle\nu, q\rangle\in\SubProb{T}\times[0, 1] \mid F(\nu)>q\}
\end{equation*}
with
\begin{align*}
  F(\nu) & := \int_{0}^{1}\nu(\hut{Q}(E, r))\ dr\\
& = \int_{0}^{1}\nu(\{t\in T\mid \basS{U}{E}{> r}\in Q(t)\})\ dr\\
& = (\nu\otimes\lambda)\bigl(\{\langle t, r\rangle\in T\times[0, 1]\mid \basS{U}{E}{> r}\in Q(t)\}\bigr)
\end{align*}
by Theorem~\ref{Choquet}; here $\nu\otimes\lambda$ is the product measure on ${\mathcal T}\otimes \Borel{[0, 1]}$ with factors $\nu$ and the Lebesgue measure $\lambda$. Then we have to show that $F$ depends measurably on $\nu$. Hence it suffices to show that $\nu\mapsto (\nu\otimes\lambda)(G)$ is a $w({\mathcal T})$-$\Borel{[0, 1]}$-measurable map for each $G\in{\mathcal T}\otimes\Borel{[0, 1]}$.

2.
Let
\begin{equation*}
  {\cal G} := \{G\in{\mathcal T}\otimes\Borel{[0, 1]}\mid \nu\mapsto (\nu\otimes\lambda)(G) \text{ is $w({\mathcal T})$-$\Borel{[0, 1]}$-measurable}\}.
\end{equation*}

Then ${\cal G}$ is a $\sigma$-algebra. This is clear from the familiar
properties of measurable maps. Moreover, ${\cal G}$ contains $D\times
B$ for $D\in{\mathcal T}, B\in\Borel{[0, 1]}$. This follows from the
definition of the weak-*-$\sigma$-algebra as the smallest
$\sigma$-algebra which renders evaluating measures measurable, and
because $(\nu\otimes\lambda)(D\times B) =
\nu(B)\cdot\lambda(B)$. Consequently, ${\cal G}$ equals ${\mathcal
  T}\otimes\Borel{[0, 1]}$.

3.
Because 
\begin{equation*}
\{\langle t, r\rangle\in T\times[0, 1]\mid \basS{U}{E}{> r}\in Q(t)\}\in {\mathcal T}\otimes\Borel{[0, 1]}
\end{equation*}
by the definition of t-measurability, we infer that $F$ is measurable, hence 
$
G'_Q(E)\in w({\mathcal T})\otimes\Borel{[0, 1]}
$
by Choquet's Theorem. But because 
$
G_{Q}(E, q) = \bigl(G'_{Q}(E)\bigr)_{q},
$
we conclude from t-measurability of $P$ that 
$
\bigl(\hut{P}\ast\hut{Q}\bigr)(E, q)\in{\mathcal S},
$
hence
$
\hut{P}\ast\hut{Q}: {\mathcal U}\times[0, 1]\to {\mathcal S}.
$ 
Monotonicity  is obvious. 
\EndProof

Let us have a look at the behavior of stochastic relations in this scenario.Given the stochastic relations $K: S \Trans T$ and $L: T \Trans U$, the \emph{convolution} (or the
\emph{Kleisli product}) $K \ast L: S \Trans U$ of $K$ and $L$ is defined
through
\begin{equation*}
  (K \ast L)(s)(V) := \int_T L(t)(V)\ K(s)(dt).
\end{equation*}
Again, binding an action $a$ to $K$, $K(s)(V)$ is interpreted as the
probability of reaching an element of $V$ upon executing action $a$ in
state $s$; binding $b$ to $L$, the probability $(K\ast L)(s)(U)$ is
interpreted as the probability for the combined action $a; b$. 

Converting a stochastic relation to a stochastic effectivity function
is compatible with convolutions:
\BeginLemma{conv-ok}
Let $K: S \Trans T$ and $L: T \Trans U$ be stochastic relations with
associated effectivity functions $P_K: S \eTrans T$ resp. $P_L:
T\eTrans U$. Then 
\begin{equation*}
  \hut{P}_K\ast\hut{P}_L = \hut{P}_{K\ast L}.
\end{equation*}
\EndLemma

\BeginProof
We obtain for $F\in{\mathcal U}, 0 \leq r \leq 1$ by expanding definitions
\begin{equation*}
  \hut{P}_{L}(F, r) = \{t \in T \mid \basS{U}{F}{>r} \in P_L(t)\}
= \{t \in T \mid L(t)(F) > r\},
\end{equation*}
thus
\begin{equation*}
  \int_0^1 \nu(\hut{P}_{L}(F, r))\ dr > q \Leftrightarrow \int_T
  L(t)(F)\ \mu(dt) > q.
\end{equation*}
Now define
\begin{equation*}
\Gamma_{P_L}(F) := \{\langle \mu, q\rangle \mid \int_0^1
\mu(\hut{P}_{L}(F, r))\ dr > q\},
\end{equation*}
then
\begin{align*}
  \hut{P}_{K\ast L}(F, q) 
& = \{s \in S \mid \basS{U}{F}{>q} \in P_{K\ast L}(s)\} \\
& = \{s \in S \mid (K\ast L)(s)(F) > q\}\\
& = \{s \in S \mid \{\mu \in \SubProb{T} \mid \int_T L(t)(F) \ \mu(dt)
> q\}\in P_K(s)\}\\
& = \{s \in S \mid (\Gamma_{P_L}(F))_q \in P_K(s)\}\\
& = (\hut{P}_{K}\ast\hut{P}_{L})(F, q)
\end{align*}
\EndProof

Hence the convolution of stochastic relations finds it counterpart in the convolution of the monotone maps which are induced by an effectivity functions. Thus stochastic effectivity functions may be used for modelling sequentiality, similar to the use of stochastic relations when modelling, e.g., dynamic logics. 
  
\subsection{Induced monotone maps}
\label{sec:induc-monot-maps}

The map $\hut{P}$ is actually obtained as a special case. Define for $P: S \eTrans T$ and for 
$
H \in \bSS{w({\mathcal T})}
$
the set
\begin{equation*}
\Im(P)(H) := \pP_P(H) := \{\langle s, q\rangle \mid H_q\in P(s)\},
\end{equation*}
as all pairs of states and numeric values for which $H$ can be achieved, then 
\begin{equation*}
\pP_P: \bSS{w({\mathcal T})}\to\bSS{{\mathcal S}}
\end{equation*}
by the definition of t-measurability. The map $\pP$ is monotone, and clearly 
\begin{equation*}
\hut{P}(D, q) = \bigl(\pP_P(\{\langle \mu, r\rangle \in \SubProb{T}\times[0, 1]\mid \mu\in\basS{T}{D}{> r}\})\bigr)_q.
\end{equation*}




This correspondence goes even a bit deeper when considering
morphisms. Let $\pP:\bSS{w({\mathcal T})}\to\bSS{{\mathcal S}}$ and
$\pP[p']: \bSS{w({\mathcal V})}\to\bSS{w({\mathcal U})}$ be monotone
maps, and take measurable maps $f: S \to U$ and $g: T \to V$. Then
$(f, g): \pP\to\pP[p']$ is said to be a \emph{morphism} iff
\begin{equation*}
\langle f(s), q\rangle \in \pP[p'](H')
\Leftrightarrow 
\langle s, q\rangle\in \pP\bigl(\InvBild{(\SubProb{g}\times id_{[0, 1]})}{H'}\bigr)
\end{equation*}
holds for all $s\in S$ and for all $q\in[0, 1]$. This condition is equivalent to saying that 
\begin{equation*}
\InvBild{(f\times id_{[0, 1]})}{\pP[p'](H')} 
=
\pP\bigl(\InvBild{(\SubProb{g}\times id_{[0, 1]})}{H'}\bigr)
\end{equation*}
holds for all $H'\in\bSS{w({\mathcal V})}$. Consequently, this diagram of maps commutes

\begin{equation*}
  \xymatrix{
\bSS{w({\mathcal T})}\ar[d]_{\pP} && \bSS{w({\mathcal V})}\ar[d]_{\pP[p']}\ar[ll]_{(\SubProb{g}\times id_{[0, 1]})^{-1}}\\
\bSS{{\mathcal S}}&&\bSS{{\mathcal U}}\ar[ll]^{(f\times id_{[0, 1]})^{-1}}
}
\end{equation*}

\medskip

The relationship between these morphisms is fairly transparent, as we
will show now. 
  
\BeginProposition{horiz-morph}
Let $P: S \eTrans T$ and $Q: U \eTrans V$ be stochastic effectivity functions, and assume $f: S \to U$ and $g: T \to V$ are measurable. Then $(f, g): P \to Q$ is a morphism iff $(f, g): \Im(P)\to\Im(Q)$ is a morphism.
\EndProposition

\BeginProof
0.
Let $\pP := \Im(P), \pP[q] := \Im(Q)$ for easier notation. We have to show
\begin{equation*}
(f, g): P \to Q \Leftrightarrow (f, g): \pP\to\pP[q].
\end{equation*}

1.
``$\Rightarrow$''
Observe that we have
\begin{equation*}
\langle \mu, q\rangle \in \InvBild{(\SubProb{g}\times id_{[0, 1]})}{H'}
\Leftrightarrow
\mu\in\InvBild{\SubProb{g}}{H'_q}
\end{equation*}
for $H'\in\bSS{w({\mathcal V})}$, thus
\begin{multline*}
\langle  s, q\rangle \in \pP\bigl(\InvBild{(\SubProb{g}\times id_{[0, 1]})}{H'}\bigr)
 \Leftrightarrow
\bigl(\InvBild{(\SubProb{g}\times id_{[0, 1]})}{H'}\bigr)_q \in P(s)\\
 \Leftrightarrow
\InvBild{\SubProb{g}}{H'_q} \in P(s) 
 \stackrel{(\dag)}{\Leftrightarrow}
H'_q \in Q(f(s))
 \Leftrightarrow
\langle f(s), q\rangle \in \pP[q](H')\\
 \Leftrightarrow
\langle  s, q\rangle \in \InvBild{(f\times id_{[0, 1]})}{\pP[q](H')}
\end{multline*}
Here $(\dag)$ uses the assumption that $(f, g)$ is a morphism for effectivity functions.

2.
``$\Leftarrow$'' 
Now let 
$
W \in Q(f(s)),
$
for $s \in S$, so that we can find $H'\in \bSS{w({\mathcal V})}$ and $q\in [0, 1]$ with $W = H'_q$ and $\langle f(s), q\rangle \in \pP[q](H')$. Because 
\begin{align*}
\langle s, q\rangle\in \InvBild{(f\times id_{[0, 1]})}{\pP[q]}
& \stackrel{(\ddag)}{\Leftrightarrow}
\langle s, q\rangle\in \pP\bigl(\InvBild{(\SubProb{g}\times id_{[0, 1]})}{H'}\bigr)\\
& \Leftrightarrow
\bigl(\InvBild{(\SubProb{g}\times id_{[0, 1]})}{H'}\bigr)_q \in P(s)\\
& \Leftrightarrow
\InvBild{\SubProb{g}}{H'_q} \in P(s)\\
& \Leftrightarrow
\InvBild{\SubProb{g}}{W} \in P(s)
\end{align*}
(using the assumption in $(\ddag)$) we find that 
$
Q(f(s)) = \Vau{g}(P(s)),
$
establishing the claim.
\EndProof

Thus morphisms for stochastic effectivity functions are in natural
bijective correspondence with morphisms for monotone maps. For plain
effectivity functions between sets, this correspondence is easily
established and silently used for the interpretation of game
logic. Adding quantitative information and working with distributions
rather than with states renders these relationships somewhat more
complicated, but exhibits a very similar structure.

\subsection{Congruences}
\label{sec:congruences}

Let $P: S \eTrans T$ be a fixed stochastic effectivity function. 

Congruences are defined as usual through morphisms and
factorization. Because we define effectivity functions between two
spaces, a congruence will have to capture properties of both spaces,
so in this general setting a congruence is a pair. The idea is that if
two elements $s, s'$ of $S$ cannot be separated through the
equivalence on $S$, then it should not be possible to separate the
portfolios of $P(s)$ and $P(s')$ through the corresponding equivalence
on $\SubProb{T}$. For expressing this adequately, we require an effectivity function on the
respective factor spaces which is compatible with the factor
structure. This leads to the following definition.

\BeginDefinition{def-congr}
A \emph{congruence for $P$} is a pair $(\alpha, \beta)$ of equivalence
relations on $S$ resp. $T$ such that there exists an effectivity
function $P_{(\alpha, \beta)}: \Faktor{S}{\alpha}\eTrans\Faktor{T}{\beta}$ which renders this diagram commutative
\begin{equation*}
\xymatrix{
S\ar[d]_P\ar[rr]^{\fMap{\alpha}} && \Faktor{S}{\alpha}\ar[d]^{P_{(\alpha, \beta)}}\\
\Vau{T}\ar[rr]_{\VauSenza{\fMap{\beta}}}  && \Vau{\Faktor{T}{\beta}}
}
\end{equation*}
If $\cG{c} = (\alpha, \beta)$ is a congruence for $P$, the effectivity function $P_{(\alpha, \beta)}$ is also denoted by $\Faktor{P}{\cG{c}}$. 
\EndDefinition

Consequently, we have 
\begin{equation*}
W \in P_{(\alpha, \beta)}(\Klasse{s}{\alpha})
\Leftrightarrow 
\InvBild{(\SubProbSenza{\fMap{\beta}})}{W}\in P(s)
\end{equation*}
for $W \in w({\Faktor{{\mathcal T}}{\beta}})$ and $s \in S$. So if $\isEquiv{s}{s'}{\alpha}$, we have in particular 
$\InvBild{(\SubProbSenza{\fMap{\beta}})}{W}\in P(s)$ iff $\InvBild{(\SubProbSenza{\fMap{\beta}})}{W}\in P(s')$, which means that $P(s)$ and $P(s')$ cannot separate those portfolios which are indistinguishable under $\SubProbSenza{\fMap{\beta}}$.

Because $\fMap{\alpha}$ is onto, $P_{(\alpha, \beta)}$ is uniquely
determined. The next proposition provides a criterion for an
equivalence relation to be a congruence. It requires the equivalence
relation $\alpha$ to be tame.

\BeginProposition{is-a-congruence}
Given a stochastic effectivity function $P: S \eTrans T$ and 
equivalence relations $\alpha$ on $S$ and $\beta$ on $T$ with $\alpha$ tame, these statements are equivalent
\begin{enumerate}
\item\label{is-a-congruence-1} $(\alpha, \beta)$ is a congruence for $P$.
\item\label{is-a-congruence-2} Whenever $\isEquiv{s}{s'}{\alpha}$, we have 
$
\InvBild{(\SubProbSenza{\fMap{\beta}})}{A} \in P(s) \text{ iff } \InvBild{(\SubProbSenza{\fMap{\beta}})}{A} \in P(s')
$
for every $A\in w(\Faktor{{\mathcal T}}{\beta})$
\end{enumerate}
\EndProposition

\BeginProof
``\labelImpl{is-a-congruence-1}{is-a-congruence-2}'': This follows immediately from the definition. 

``\labelImpl{is-a-congruence-2}{is-a-congruence-1}'': Define for $s
\in S$ 
\begin{equation*}
P_{(\alpha, \beta)}(\Klasse{s}{\alpha}) := \{A \in w(\Faktor{{\mathcal T}}{\beta}) \mid 
\InvBild{\SubProb{\fMap{\beta}}}{A} \in P(s)\},
\end{equation*}
then $P_{(\alpha, \beta)}$ is well defined by the assumption, and it
is clear that $P_{(\alpha, \beta)}(\Klasse{s}{\alpha})$ is an upward
closed set of subsets of  $w(\Faktor{{\mathcal T}}{\beta})$ for each $s \in
S$. It remains to show that $P_{(\alpha, \beta)}$ is a stochastic
effectivity function, i.e., that $P_{(\alpha, \beta)}$ is
t-measurable. 

In fact, let 
$
H \in \bSS{w(\Faktor{{\mathcal T}}{\beta})}
$
be a test set, and put 
\begin{equation*}
  Y :=   \{\langle\Klasse{s}{\alpha}, q\rangle \mid H_q \in
  P_{(\alpha, \beta)}(\Klasse{s}{\alpha})\} 
=   \Bild{(\fMap{\alpha}\times id_{[0, 1]})}{Z}
\end{equation*}
with
\begin{equation*}
Z :=  \{\langle s, q\rangle \mid \InvBild{\SubProb{\fMap{\beta}}}{H_q}\in P(s)\}
\end{equation*}
as its inverse image under $\SubProb{\fMap{\alpha}}\times id_{[0,
  1]}.
$
Then
$Z$ is $(\alpha\times \Delta)$-invariant. By Corollary~\ref{cor-is-tame}
it is enough to show that $Z$ is a member of 
$\Sigma_{\alpha}\otimes\Borel{[0, 1]}$. Then
$
Y \in \Faktor{{\mathcal S}}{\alpha}\otimes[0, 1]
$
will follow. 

Because $P$ is t-measurable, we infer 
$
Z \in\bSS{{\mathcal S}}.
$
Since $Z$ is $(\alpha\times\Delta)$-invariant, we conclude from
$
\InvBild{\SubProb{\fMap{\beta}}}{H_q} =
  \bigl(\InvBild{(\SubProb{\fMap{\beta}}\times id_{[0, 1]})}{H}\bigr)_q 
$
and
$
\InvBild{(\SubProb{\fMap{\beta}}\times id_{[0, 1]})}{H} \in
\bSS{w({\mathcal T})}
$
that 
$
Z \in \Sigma_{\alpha\times\Delta}.
$
The latter $\sigma$-algebra is equal to 
$
\bSS{\Sigma_{\alpha}}
$
by Lemma~\ref{is-tame}, because $\alpha$ is tame. 
\EndProof


As a consequence, the kernel of a morphism $(f, g)$ is a congruence,
provided $\Kern{f}$ is tame, and provided $g$ is final. The proof uses the observation
formulated in Corollary~\ref{isomorphism-f-final} that equivalence
relations induced by the kernel of a final and surjective measurable
map preserve the $\sigma$-algebras on which they operate.

\BeginProposition{kernel-morph} 
Let  $(f, g): P \to Q$ be a morphism
for the stochastic effectivity functions $P: S \eTrans T$ and $Q: U
\eTrans V$. If $f \times id_{[0, 1]}$ and $g$ are final, then
$(\Kern{f}, \Kern{g})$ is a congruence for $P$. 
\EndProposition

\BeginProof
1.  If $f\times id_{[0, 1]}$ is final, then $\Kern{f\times id_{[0,
    1]}} = \Kern{f}\times\Delta$ is tame by
Lemma~\ref{kerf-is-exact}. Hence it remains to show that
condition~\ref{is-a-congruence-2} in Proposition~\ref{is-a-congruence}
is satisfied.

2.
Decompose $g = \widetilde{g}\circ \fMap{\Kern{g}}$ with $\widetilde{g}: \Faktor{T}{\Kern{g}}\to V$ (see the proof of Lemma~\ref{kerf-is-exact}), and put 
\begin{equation*}
  \mathcal{Z} := \{A \in w(\Faktor{{\mathcal T}}{\Kern{g}}))\mid A = \InvBild{(\SubProbSenza{\widetilde{g}})}{A_1}\text{ for some } A_1\in w({\mathcal V})\}.
\end{equation*}
Then $\mathcal{Z}$ is a $\sigma$-algebra; we show that 
$
\mathcal{Z} = w(\Faktor{{\mathcal T}}{\Kern{g}}).
$
Let $G \in \Faktor{{\mathcal T}}{\Kern{g}}$, then finality of $g$ implies that we find $G_1\in{\mathcal V}$ with 
$
G = \InvBild{\widetilde{g}}{G_1},
$
as in the proof of Lemma~\ref{kerf-is-exact}, thus
\begin{equation*}
  \basS{\Faktor{T}{\Kern{g}}}{G}{> q}
 = \basS{\Faktor{T}{\Kern{g}}}{\InvBild{\widetilde{g}}{G_1}}{> q} 
 = \InvBild{\SubProb{\widetilde{g}}}{\basS{U}{G_1}{> q}}
\end{equation*}
Consequently, 
\begin{equation*}
  \{\basS{\Faktor{T}{\Kern{g}}}{G}{> q} \mid G \in \Faktor{{\mathcal T}}{\Kern{g}}\} \subseteq \mathcal{Z}.
\end{equation*}
Since the former set generates $w(\Faktor{{\mathcal T}}{\Kern{g}})$, the claim follows. 

3.
Now let $B \in w(\Faktor{{\mathcal T}}{\Kern{g}})$, hence we find $B_1\in w({\mathcal T})$ with 
$
\InvBild{\SubProb{\fMap{\Kern{g}}}}{B} = \InvBild{\SubProb{g}}{B_1},
$
thus we obtain for $\langle s, s'\rangle \in \Kern{f}$ 
\begin{align*}
  \InvBild{\SubProb{\fMap{\Kern{g}}}}{B} \in P(s)
& \Leftrightarrow \InvBild{\SubProb{g}}{B_1} \in P(s) \\
& \Leftrightarrow B_1 \in Q(f(s)) = Q(f(s')) \\
& \Leftrightarrow \InvBild{\SubProb{\fMap{\Kern{g}}}}{B} \in P(s'),
\end{align*}
because $(f, g)$ is a morphism. 
\EndProof

A morphism $(f, g): P \to Q$ is called \emph{strong} iff $f \times
id_{[0, 1]}$ and $g$ are final; a congruence $(\alpha, \beta)$ is
called \emph{tame} iff $\alpha$ is a tame equivalence relation. Thus
we have shown that the kernel of a strong morphism is a tame
congruence, and vice versa. 

The correspondence between strong morphisms and tame equivalence
relations will turn out to be fairly tight, as we will see when
investigating logical and behavioral equivalence of stochastic
effectivity functions.


\section{Logical and Behavioral Equivalence}
\label{sec:expressivity}

Interpreting a logic, the question of expressivity of the underlying
models is important. For example, logical equivalence requires that we
find for each state in one model another state which satisfies exactly
the same formulas; through factoring we then will be able under some
circumstances to build smaller models with the same expressive
power. We are in a position now to characterize logical
resp. behavioral equivalence through morphisms and congruences, and to
relate these notion of expressivity in purely algebraic terms, i.e.,
without reference to an underlying logic. Behavioral equivalence is
expressed through a co-span of surjective morphisms, and logical
equivalence is expressed through isomorphic factor
spaces. Investigating the relationship between the two, we show that
logically equivalent effectivity functions are behaviorally
equivalent, and that the converse holds as well, provided we assume
that the morphisms are strong.

Given two stochastic effectivity functions $P: S \eTrans T$ and $Q: U
\eTrans V$, call $P$ and $Q$ \emph{logically equivalent} iff there
exist tame congruences $\cG{c}$ for $P$ and $\cG{d}$ for $Q$ such that
$\Faktor{P}{\cG{c}}$ and $\Faktor{Q}{\cG{d}}$ are isomorphic. The name
derives from an observation for Kripke models in modal logics: two
states are called equivalent iff they have the same theory, i.e.,
accept exactly the same formulas, and two Kripke models are called
logically equivalent iff given a state in one model, there exists a
state in the other one with the same theory. Then it can be shown for
stochastic Kripke models over analytic spaces in a fairly general,
coalgebraic context that the corresponding factor models are
isomorphic~\cite[Corollary 4.8, Theorem 6.17]{EED-CS-Survey}.

Similarly, call $P$ and $Q$ \emph{behaviorally equivalent} iff there
exists a mediating function $M: X \eTrans Y$ and strong surjective
morphisms $(f, g): P \to M$ and $(k, \ell): Q \to M$. Thus we obtain
the familiar diagram
\begin{equation}
\label{eq:4} \xymatrix{
S \ar[rr]^{f}\ar[d]_P&& X\ar[d]_M && U\ar[ll]_{k}\ar[d]^Q\\
\Vau{T}\ar[rr]_{\VauSenza{g}} && \Vau{Y} && \Vau{V}\ar[ll]^{\VauSenza{\ell}}
}
\end{equation}
This diagram translates into
\begin{align}
\label{eq:2}  D \in M(f(s)) & \Leftrightarrow \InvBild{(\SubProbSenza{g})}{D} \in P(s)\\
\label{eq:3}  D \in M(k(u)) & \Leftrightarrow \InvBild{(\SubProbSenza{\ell})}{D} \in Q(u)
\end{align}
for $D\subseteq\SubProb{Y}$ measurable and $s\in S, u \in U$. 

The correspondence between the use of tame equivalence relations for
logical equivalence and of strong morphisms for behavioral equivalence
is noteworthy. From a technical point of view, tame relations are
necessary for constructing the factor model. This suggests the use of
strong morphisms, because then the corresponding kernels form a
congruence (see Proposition~\ref{is-a-congruence}). Thus we require a
strong morphism in order to factor through the kernels, which in turn
enables us to compare factor models.
 
We fix for the sequel the stochastic effectivity functions $P: S \eTrans T$ and $Q: U \eTrans V$. 

\BeginProposition{log-impl-beh}
If $P$ and $Q$ are logically equivalent, they are behaviorally equivalent.
\EndProposition

\BeginProof
If $\alpha$ is a tame equivalence relation, then $\fMap{\alpha}\times id_{[0, 1]}$ is final. 
\EndProof

\makeatletter
\def\@kl#1#2{\Klasse{#1}{{#2}}}
\newcommand{\kls}[1][s]{\@kl{#1}{f}}
\newcommand{\klu}[1][u]{\@kl{#1}{k}}
\newcommand{\klt}[1][t]{\@kl{#1}{g}}
\newcommand{\klv}[1][v]{\@kl{#1}{\ell}}
\def\@af#1#2{\Faktor{#1}{{#2}}}
\newcommand{\aSf}{\@af{S}{f}}
\newcommand{\aTg}{\@af{T}{g}}
\newcommand{\aUk}{\@af{U}{k}}
\newcommand{\aVl}{\@af{V}{\ell}}
\newcommand{\aVll}{w(\@af{{\mathcal V}}{\ell})}
\newcommand{\aPfg}{\@af{P}{f, g}}
\newcommand{\aQkl}{\@af{Q}{k, \ell}}
\makeatother

In order to show that behaviorally equivalent effectivity functions
are logically equivalent, fix $M: X \eTrans Y$ and the morphisms
according to diagram~(\ref{eq:4}) with measurable spaces $X$ and $Y$;
to make notation not heavier than it is, we do without an explicit
name for the $\sigma$-algebras on $X$ resp. $Y$. Recall that we simplify notations by writing, e.g., $\kls$ rather that $\Klasse{s}{\Kern{f}}$, similarly for $\Sigma_{f}$. 
\begin{align*}
  \gamma & := \{\langle \kls, \klu\rangle \mid s \in S, u \in U, f(s) = k(u)\}\\
  \delta & := \{\langle \klt, \klv\rangle \mid t \in T, v \in V, g(t) = \ell(v)\}\\
\end{align*}

Because the contributing maps are onto, $\gamma$ and $\delta$ are the
graphs of bijective maps; this is shown exactly as in the proof
of~\cite[Lemma 2.6.10]{EED-CoalgLogic-Book}.  For simplicity, the maps
proper are called $\delta$ and $\gamma$ as well.

\BeginLemma{are-meas}
$
  \gamma:   \Faktor{S}{\Kern{f}}\to\Faktor{U}{\Kern{k}}
$
and
$
  \delta:   \Faktor{T}{\Kern{g}}\to\Faktor{V}{\Kern{\ell}}  
$
are Borel isomorphisms.
\EndLemma

\BeginProof
We show that 
$
\InvBild{\gamma}{E} \in \Faktor{{\mathcal S}}{\Kern{f}}
$
for $E \in \Faktor{{\mathcal U}}{\Kern{k}}$. Because by construction
$
\InvBild{\fMap{k}}{E}\in\Sigma_{k},
$
and because $k$ is final, we find a measurable subset $H\subseteq X$ with 
$
\InvBild{\fMap{k}}{E} = \InvBild{k}{H}.
$
Put 
$
G := \InvBild{f}{H},
$
then 
$
G \in\Sigma_{f}.
$
Thus
$
G_0 := \Bild{\fMap{f}}{G}\in\Faktor{{\mathcal S}}{\Kern{f}}.
$
So we are done, provided we can show that $\InvBild{\gamma}{E} = G_0$ holds. 

In fact, let $ \kls\in\InvBild{\gamma}{E} $ with $ \klu:=\gamma(\kls) \in
E, $ thus $ f(s) = k(u).  $ Hence $ u \in \InvBild{k}{H}, $ which implies $ f(s) \in H, $ so that $ s \in \InvBild{f}{H} =
G, $ which in turn means $ \kls\in G_0, $ so that $ \InvBild{\gamma}{E}\subseteq
G_0.  $ On the other hand, if $\kls\in G_{0}$, we know that $s\in G$.
Hence $f(s)\in H$, so that we can find $u\in U$ with $f(s) =
k(u)$. Since $k(u)\in H$, we know
$u\in\InvBild{k}{H}=\InvBild{\fMap{k}}{E}$, but this means $\klu =
\gamma(\kls)\in E$, establishing the other inclusion.

The claim for $\delta$ is established in exactly the same way. 
\EndProof

Because $(f, g): P \eTrans M$ is a strong morphism, its kernel
$\Kern{f, g} := (\Kern{f}, \Kern{g})$ is a congruence by
Proposition~\ref{kernel-morph}, similarly for $(k, \ell): Q \eTrans
M$. Hence the factor functions $ \Faktor{P}{\Kern{f, g}}:
\Faktor{S}{\Kern{f}}\eTrans\Faktor{T}{\Kern{g}} $ and $
\Faktor{Q}{\Kern{k, \ell}}:
\Faktor{U}{\Kern{k}}\eTrans\Faktor{V}{\Kern{\ell}} $ exist.

It turns out that these functions yield isomorphisms.

\BeginProposition{are-isomorphic}
$(\gamma, \delta): \Faktor{P}{\Kern{f, g}} \to \Faktor{Q}{\Kern{k, \ell}}$ is an isomorphism.
\EndProposition

\BeginProof
0.  We know from Lemma~\ref{are-meas} that both $\gamma$ and $\delta$
are Borel isomorphisms, so we have to show that they are compatible
with the structure of the effectivity functions. We show that this
diagram commutes

\begin{equation*}
  \xymatrix{
\Faktor{S}{\Kern{f}}\ar[d]_{\Faktor{P}{\Kern{f, g}}} \ar[rr]^{\gamma}&& \Faktor{U}{\Kern{k}}\ar[d]^{\Faktor{Q}{\Kern{k, \ell}}}\\
\Vau{\Faktor{T}{\Kern{g}}} \ar[rr]_{\VauSenza{\delta}}&&\Vau{\Faktor{V}{\Kern{\ell}}}
}
\end{equation*}
Interchanging the r\^oles of $\gamma$ and $\delta$ will then establish the result. 

1.
Fix $G \in \aVll$ and $s \in S$, we show that 
\begin{equation*}
G \in (\aQkl)(\gamma(\kls)) 
\Leftrightarrow
\InvBild{(\SubProbSenza{\delta})}{G} \in (\aPfg)(\kls).
\end{equation*}
From Lemma~\ref{hilfslemma} we infer that we can find for $G$ a measurable set $H \subseteq\SubProb{Y}$ such that
\begin{align}
\label{sec:expressivity-1} \InvBild{\SubProb{\delta\circ\fMap{g}}}{G} & = \InvBild{(\SubProbSenza{g})}{H}\\
\label{sec:expressivity-2} \InvBild{(\SubProbSenza{\fMap{\ell}})}{G} & = \InvBild{(\SubProbSenza{\ell})}{H}
\end{align}
Given $s \in S$, we find $u\in U$ such that $f(s) = k(u)$. Then we have
 \begin{multline*}
   \InvBild{(\SubProbSenza{\delta})}{G}\in(\aPfg)(\kls)
 \Leftrightarrow \InvBild{(\SubProbSenza{\fMap{g}})}{\InvBild{(\SubProbSenza{\delta})}{G}}\in P(s)\\
  \stackrel{(\ref{sec:expressivity-1})}{\Leftrightarrow}
\InvBild{(\SubProbSenza{g})}{H}\in P(s)
 \stackrel{(\ref{eq:2})}{\Leftrightarrow}
H \in M(f(s)) = M(k(u))
 \stackrel{(\ref{eq:3})}{\Leftrightarrow}
\InvBild{(\SubProbSenza{\ell})}{H}\in Q(u) \\
 \stackrel{(\ref{sec:expressivity-2})}{\Leftrightarrow}
\InvBild{(\SubProbSenza{\fMap{\ell}})}{G} \in Q(u) 
 \Leftrightarrow
G \in (\aQkl)(\gamma(\kls))
 \end{multline*}
This was to be shown.
\EndProof

We have delayed, however, the proof of an auxiliary statement.

\BeginLemma{hilfslemma}
For every $G\in \aVll$ there exists a measurable set $H \subseteq\SubProb{Y}$ such that
\begin{equation} 
\label{eq:10}\InvBild{\SubProb{\delta\circ\fMap{g}}}{G}  = \InvBild{(\SubProbSenza{g})}{H} \text{ and }
 \InvBild{(\SubProbSenza{\fMap{\ell}})}{G}  = \InvBild{(\SubProbSenza{\ell})}{H}
\end{equation}
\EndLemma
\BeginProof
1.
Let $\mathcal{Z}$ be the set of all $G\in \aVll$ such that the assertion is true. Then $\mathcal{Z}$ is a $\sigma$-algebra, so that it is sufficient to demonstrate that the statement is true for a generator
$
G = \basS{\Faktor{V}{\Kern{\ell}}}{A}{> q}
$
with 
$
A \in \Faktor{{\mathcal V}}{\Kern{\ell}}
$
and $0 \leq q \leq 1$.

2.
Because $A \in \Faktor{{\mathcal V}}{\Kern{\ell}}$, we know that 
$
A = \Bild{\fMap{\ell}}{A_0}
$
for some 
$
A_0 \in\Sigma_{\ell}.
$ 
Thus 
$
A_0 = \InvBild{\ell}{H_0}
$
for some measurable set $H_0 \subseteq Y$. We claim that 
$
H := \basS{Y}{H_0}{> q}
$
is the set we are looking for in~(\ref{eq:10}).

3.
First we note that
$
A_0 = \InvBild{\fMap{\ell}}{A},
$
hence
\begin{multline*}
  \InvBild{(\SubProbSenza{\ell})}{H}
= \InvBild{(\SubProbSenza{\ell})}{\basS{Y}{H_0}{> q}}
= \basS{V}{\InvBild{\ell}{H_0}}{> q}
= \basS{V}{A_0}{> q}\\
= \basS{V}{\InvBild{\fMap{\ell}}{A}}{> q}
= \InvBild{(\SubProbSenza{\fMap{\ell}})}{\basS{{\Faktor{V}{\ell}}}{A}{> q}}
= \InvBild{(\SubProbSenza{\fMap{\ell}})}{G}.
\end{multline*}

4.
Then we claim that 
\begin{equation}
\label{bild-eq}
t\in \InvBild{g}{H_{0}}
\Leftrightarrow
\delta(\klt) \in \Bild{\fMap{\ell}}{\InvBild{\ell}{H_0}}
\end{equation}
Assume that $g(t)\in H_{0}$, and take $v \in V$ with $g(t) = \ell(v)$, thus
$
\delta(\klt) = \klv,
$
hence 
$
\klv\in\Bild{\fMap{\ell}}{\InvBild{\ell}{H_0}}.
$
On the other hand, if 
$
\delta(\klt) \in \Bild{\fMap{\ell}}{\InvBild{\ell}{H_0}},
$ 
there exists $v\in V$ such that $g(t) = \ell(v)$, so that 
$
\ell(v) \in H_0,
$
which implies 
$
t \in \InvBild{g}{H_0}.
$

Thus we obtain
{\def\xxX{\Faktor{V}{\Kern{f}}}
\begin{multline*}
  \InvBild{(\SubProbSenza{g})}{H}
 = \InvBild{(\SubProbSenza{g})}{\basS{Y}{H_0}{> q}}
 = \basS{T}{\InvBild{g}{H_{0}}}{>q}\\
 \stackrel{(\ref{bild-eq})}{=} \basS{T}{\InvBild{\fMap{g}}{\InvBild{\delta}{\Bild{\fMap{\ell}}{\InvBild{\ell}{H_{0}}}}}}{>q}
 = \Bild{\SubProbSenza{(\delta\circ \fMap{g})}}{\basS{\xxX}{\Bild{\fMap{\ell}}{\InvBild{\ell}{H_{0}}}}{>q}}\\
 = \Bild{\SubProbSenza{(\delta\circ \fMap{g})}}{\basS{\xxX}{\Bild{\fMap{\ell}}{A_{0}}}{>q}}
 = \Bild{\SubProbSenza{(\delta\circ \fMap{g})}}{\basS{\xxX}{A}{>q}}
 = \Bild{\SubProbSenza{(\delta\circ \fMap{g})}}{G}
\end{multline*}
}
Consequently, $H$ is the set we are looking for. 

5. 
This implies
\begin{equation*}
  \{\basS{\Faktor{V}{\Kern{\ell}}}{A}{> q} \mid A \in \Faktor{{\mathcal V}}{\Kern{\ell}}, 0 \leq
  q \leq 1\} \subseteq \mathcal{Z}, 
\end{equation*}
so that 
$
\mathcal{Z}=\aVll.
$
\EndProof

\medskip

Summarizing, we have shown
\BeginProposition{beh-impl-log}
Behaviorally equivalent stochastic effectivity functions are logically equivalent.
\QED
\EndProposition

Thus we observe a close correspondence of behavioral and logical
equivalence, which might be compared to a similar result obtained
in~\cite{EED-CongBisim} for stochastic relations. There these
equivalences are compared with each other and with bisimilarity, and
under fairly strong topological assumptions it was shown that a result
similar to Proposition~\ref{beh-impl-log} can be obtained. We do not
impose topological assumptions in the present paper, but rather
require the morphisms to be strong, and the equivalence relations
to be tame, so there is a trade off among assumptions. On the other
hand, Proposition~\ref{pointed} tells us 
that stochastic effectivity functions are strictly more general than
stochastic relations. 


\section{Conclusion}
\label{sec:conclusion}

The algebraic properties of stochastic effectivity functions have
been studied, and we have investigated the question of expressivity of
these functions. Logical and behavioral equivalence have been related
to each other; this has been done without recourse to an underlying
logic. 

Bisimilarity is usually a companion to logical and to behavioral
equivalence. Call two stochastic effectivity functions $P$ and $Q$
\emph{bisimilar} iff there is a mediating function $M$ to which $P$ and $Q$
are related through a span $P \leftarrow M \rightarrow Q$ of morphisms. This is the
coalgebraic definition of bisimilarity. In the context of effectivity
functions for games, a relational definition is used, quite close to
Milner's original one~\cite{Blackburn-Rijke-Venema,Pauly-Parikh}. It
was shown, however, that the relational and the coalgebraic one are
equivalent~\cite[Proposition 1.142]{EED-Categs}, provided the
relation's projections are surjective (see also~\cite{HK04}).

While it is possible to relate logical and behavioral equivalence
algebraically for general measurable spaces, this seems to be more
involved for the case of bisimilarity. In a comparable situation for
stochastic relations, it was shown that bisimilar relations are
logically equivalent under a compactness assumption, the converse was
established for Polish spaces through a selection
argument~\cite{EED-CongBisim}; see~\cite{EED-CS-Survey} for a broader survey. 

Related questions remain to be looked at carefully as well, among them
the relationship to non-deterministic labeled Markov processes, in
particular the question to determine under which conditions such a
process is generated by a stochastic effectivity function. Certainly,
t-measurability plays a crucial r\^ole;~\cite{EED+PST} discusses some
of these issues, introduces subsystems, and proposes a partial
solution to the problem of bisimilar effectivity functions, as
expected, under some topological conditions.

\paragraph{Acknowledgements}
\label{sec:acknowledgements}

The referees both made some very helpful suggestions to clarify the paper's
representation and content. Many long discussions with Pedro Sánchez
Terraf helped me to improve my understanding of NLMPs and their often
intricate relationship to stochastic effectivity functions. All this
is appreciated.


\end{document}